\documentclass{article}

\usepackage{graphicx}
\usepackage{amsmath,amssymb}

\def\ds{\displaystyle}
\def\bea{\begin{array}{c}}
\def\ea{\end{array}}
\def\be{\begin{equation}\bea\ds}
\def\ee{\ea\end{equation}}
\def\bee{\begin{equation}\begin{array}{rcl}\ds}
\def\eee{\end{array}\end{equation}}

\def\nn{\nonumber}

\def\Tr{{\rm Tr}\,}

\def\MS{\mathfrak{S}}

\def\be{\begin{eqnarray}}
\def\ee{\end{eqnarray}}
\def\nn{\nonumber}

\def\p{\partial}

\def\Tr{{\rm Tr}\,}

\def\l[{\phantom.[}

\hoffset= - 1.0in         % switch off for draft style
\begin{document}

\title{{\bf {On skew tau-functions in higher spin theory
}\vspace{.2cm}}
\author{{\bf D. Melnikov$^{a,b}$}, \ {\bf A. Mironov$^{a,c,d,e}$}, \ {\bf A. Morozov$^{a,d,e}$}}
\date{ }
}

\maketitle

\vspace{-5.5cm}

\begin{center}
\hfill FIAN/TD-02/16\\
\hfill IITP/TH-02/16\\
\hfill ITEP/TH-03/16\\
\end{center}

\vspace{4.2cm}

\begin{center}
$^a$ {\small {\it ITEP, Moscow 117218, Russia}}\\
$^b$ {\small {\it International Institute of Physics, UFRN
Av. Odilon G. de Lima 1722, Natal 59078-400, Brazil}}\\
$^c$ {\small {\it Lebedev Physics Institute, Moscow 119991, Russia}}\\
$^d$ {\small {\it National Research Nuclear University MEPhI, Moscow 115409, Russia }}\\
$^e$ {\small {\it Institute for Information Transmission Problems, Moscow 127994, Russia}}
\end{center}

\vspace{.5cm}

\begin{abstract}
Recent studies of higher spin theory in three dimensions
concentrate on Wilson loops in Chern-Simons theory,
which in the classical limit reduce to peculiar corner
matrix elements between the highest and lowest weight states
in a given representation of $SL(N)$.
Despite these "skew" tau-functions can seem very different
from conventional ones, which are the matrix elements
between the two highest weight states,
they also satisfy the Toda recursion
between different fundamental representations.
Moreover, in the most popular examples they possess simple
representations in terms of matrix models and Schur functions.
We provide a brief introduction to this new interesting
field, which, after quantization, can serve as
an additional bridge between knot and integrability theories.
\end{abstract}

%\bigskip

\bigskip

\section*{Introduction}

Conformal symmetry \cite{CFT} and its $W_N$ and Kac-Moody extensions in $2d$ are big enough to unambiguously
define all correlation functions and essentially reduce them to those of free fields: the only remaining freedom is the choice of evolution operators
and projection to various subsectors, closed under operator expansion.
In higher dimensions, the same role is presumably played by "higher-spin symmetries" \cite{hsa},
though reduction to free fields is not yet fully described  and understood.
The simplest is the situation in $3d$, where the higher spin theories are actually
identified with Chern-Simons theory \cite{CS}, which was much technically developed
in recent years.
An old conjecture \cite{Wit} identifies quantum $3d$ gravity with $SL(2,R)\times SL(2,R)$
Chern-Simons ($SL(2,C)$ in the Euclidean case), and $SL(N,R)\times SL(N,R)$ is considered as a natural generalization
towards higher spins \cite{CSgN}.
Specifics of $3d$ is the lack of propagating gravitons, thus the corresponding
(Chern-Simons) sector of the theory is topological
and essentially reduces to the study of $2d$ conformal blocks and their modular transformations.
Analytical continuation in $N$, needed for revealing the true algebra of $3d$ higher spin theory \cite{hsa},
is a standard procedure in Chern-Simons case, where observables (knot polynomials \cite{knotpols})\footnote{Note that, in the case of $SL(N)$, one needs non-compact knot invariants, see a recent review in \cite{GMM}.}
are usually functions of parameters $A$ and $q$, and specialization to $SL(N)$ is provided by putting $A=q^N$. Here $q=e^{2\pi i\over \kappa +N}$, where $\kappa$ is the Chern-Simons coupling constant. Moreover, if one considers the knot superpolynomial \cite{super} within the refined Chern-Simons \cite{refinedCS}, one has to make, upon specialization, an additional reduction in order to get categorification of the polynomial for a concrete group \cite{KhR}. This is much similar to the way how the higher spin algebra reduces at integer values of $N$.
Extension from $SL(N)$ to other algebras, orthogonal, symplectic and exceptional, is also
possible and one can even search for "universality" \cite{vogeluniv},
which substitutes two parameters $q,A$  by a triple $u,v,w$
and provides common description of the "$E_8$-sector" of representation theory
for all simple Lie algebras at once.

\bigskip

Developments in the field follow the standard logic:
\be
\begin{array}{ccc}
\text{geometry} & \longrightarrow & \text{algebra}\\
&&\\
&& \downarrow\\
&&\\
\text{generalized geometry} & \longleftarrow & \text{change of the algebra}
\end{array}
\label{GAAG}
\ee
The basic step in the first line is the substitution of metric $g_{\mu\nu}$ by the
$SL(2)$-valued dreibein $e_\mu$ and spin-connection $\omega_\mu$
by the standard rule $g_{\mu\nu} = \Tr e_\mu e_\nu$
and observation that Einstein metrics, satisfying $R_{\mu\nu}=\frac{1}{\ell^2}\, g_{\mu\nu}$,
are associated with flat $SL(2)\times SL(2)$  connections
$A = \frac{1}{\ell}\, e + \omega, \ \ \bar A = \frac{1}{\ell}\, e - \omega$.
This allows one to embed classical general relativity (with the $\Lambda$-term $\ell^{-2}$)
in $3d$ into Chern-Simons theory.
However, for $N>2$ this embedding is not gauge invariant, and the same flat connection in
different gauges can provide different classical metrics, what is a rich source
of speculations about the meaning of classical gravity and enhancement of symmetries
at quantum level.
In practice, one just picks up various flat connections and considers
associated Wilson lines, which can be open.

In this way a number of interesting checks was already performed.
Mainly one can compare in various examples
the geodesic lengths with the open Wilson line integrals and
identify
\be
\text{a function}\Big(-\text{geodesic length}\Big) = \text{matrix element}
\label{geodlength=mamo}
\ee
In known examples the "function" is hyperbolic cosine and the
"matrix element" is a trace of product of P-exponentials from two
different $SL(N)$-constituents of the gauge group.
In  the simplest cases of essentially constant connections
the chain of relations is even richer:
\be
e^L = \exp\Big(\text{entanglement entropy}\Big)
= \text{classical Wilson line} =
\ \left<-\text{hw}_\rho|e^{wT_+ + \bar w T_-}|\text{hw}_\rho\right>
\label{adscft}
\ee
where {\bf geodesic and the Wilson line end at the boundary of AdS space}. This is since emergency of the entanglement entropy is here due to the second ingredient, to the AdS/CFT correspondence. In contrast with the gravity/Chern-Simons correspondence, this one requires dealing with Wilson lines ending at the boundary.
The entanglement entropy is defined by the replica method,
as a limit of the Renyi entropies of CFT at the boundary $\partial\Big({{\rm AdS}_3}\Big)=S^2$
calculated in the standard way \cite{Zam},
when conformal theory on covering of the Riemann sphere is identified
with the one on $S^2$ with insertion of ramification operators.
Matrix element is that of the $SL(2)$ generators within $SL(N)$
between the highest- and the lowest-vector states of the Weyl representation $\rho$,
which is distinguished because
it is uncharged w.r.t. the generators of $SL(N)$ beyond the principally embedded
gravity-subgroup $SL(2)\in SL(N)$.
Beyond-the-geometry deformation (i.e. the one from gravity to "higher spin" theory)
appears when the group element $G$ does not belong to the $SL(2)$ subalgebra of $SL(N)$.
To get relation to higher spin theory one needs the answers with full
dependence on $N$ to allow for the complete infinite higher spin algebra.

More generally, one can consider for closed Wilson lines the character
\be
\chi_R(G)= \Tr_R G
\ee
and for open lines the matrix elements
\be
{\cal G}_{\pm R} = \left<\pm\text{hw}_R|G|\text{hw}_R\right>
= \prod_{k=1}^{N-1} \left<\pm\text{hw}_{\omega_k}|G|\text{hw}_{\omega_k}\right>^{r_k}
= \prod_{k=1}^{N-1} {\cal G}_{\pm k}^{r_k}
\label{clafactor}
\ee
between the highest and/or lowest states of {\it arbitrary} representation
$R=[r_1,r_2,r_3,\ldots]$ with the weight
$w_R= \sum_k d_k\omega_k$ and $d_k=r_k-r_{k+1}$.\footnote{
{\bf NB:} When one studies decompositions of representation products
like in knot theory applications \cite{MMMkn2},
one and the same representation can appear with non-trivial multiplicity;
then factorization is true only in a special basis, where the highest weights
are products of the fundamental ones.
Still, such a basis always exists, and this facilitates study of the
Racah matrices {\it a la} \cite{MMMSracah}.
}
Here $\omega_k$ are the fundamental weights and for the Weyl representation
$\rho = [N-1,N-2,\ldots, 1]$ of $SL(N)$ the weight is
$w_\rho = \frac{1}{2}\sum_{\Delta>0} \alpha_\Delta = \sum_k\omega_k$
with unit coefficients.
These objects are usually studied in the theory of $\tau$-functions,
and this explains why relatively explicit formulas are often available
and analytical continuations are often straightforward .

Quantum counterparts of these quantities, i.e. dependence on the parameter $q$ (related to the Chern-Simons coupling) can emerge in different ways.

Characters are naturally lifted to link   polynomials,
which depend on an extra parameter $q\neq 1$,
and they are also analytically continued in $N$.
More important, they depend non-trivially (in an essentially non-Abelian way)
on representations of other link components.
The open Wilson lines can sometime be related with link polynomials, when one link
component is considered as a "source", which imposes non-trivial boundary conditions
for the "probe" component.
An additional subject is interpretation of the matrix elements as a classical
(large-$c$ and large-dimensions) limit of the conformal blocks,
which implies a quantization related to the link invariant one.

In fact, link polynomials are natural objects from quantum group theory.
Therefore a possible task for future considerations
is to lift the series expansions for ${\cal G}_R$
(in powers of deviation of $G$ from exactly solvable cases
like the principally embedded $SL(2)\in SL(N)$ or low-triangular matrices)
to the quantum group level,
where expansions can probably be related to those of the colored link polynomials
$H_{[r_1r_2]}$ with $r_2\ll r_1$ in powers of $r_2/r_1$.
Moreover, the group element $G$ for quantum groups defined to preserve the factorization property
(\ref{clafactor}) is necessarily operator valued, hence such should be its
matrix elements.
Taken literally, one can make this way the entanglement entropy operator valued as well.

\bigskip

The story naturally decomposes into three pieces.

The first one reduces a problem of calculating geodesic lengths
in the Chern-Simons reformulation and entanglement entropies
to certain matrix elements representing classical Wilson lines,
what provides a formulation, which does not make much difference
between gravity and high spin theories.

The second is evaluation of these matrix elements by techniques
of integrability theory.

The third are speculations about generalization from classical
to quantum level.

\bigskip

Accordingly,
in section \ref{phys} we remind, what kind of group theory quantities are needed to
represent physical observables (open Wilson lines, geodesic lengths and entropies) in
$3d$ gravity and its higher spin extensions.

In sections \ref{math}-\ref{doubleskew} we demonstrate that these are exactly the quantities
studied in the theory of integrable systems, which allows one to put the recent
calculations of \cite{HKP} into this general setting.

Among other things, this sheds some additional light on the problem with
analytical continuation in $N$ encountered in \cite{HKP}.
We briefly discuss this issue  in section \ref{con},
with emphasize on the peculiarities of {\it forced} integrable hierarchies.

In fact, in section \ref{con} we discuss various quantization/deformation ideas,
which are implied by the formalism, but are not always easy to interpret
in physical terms.

%\newpage

\section{Physical background: three-dimensional gravity and Chern-Simons theory
\label{phys}}

In this section we discuss physical aspects of the three-dimensional gravity/higher spin theory in the AdS space and its relation to Chern-Simons theory. Our goal is to summarize relevant physical observables.

\subsection{Generalities}

It was originally proposed in~\cite{Wit} that the pure gravity in ${\rm AdS}_3$ is classically equivalent
to the $SL(2,R)\times SL(2,R)$ Chern-Simons theory.
Namely, a pair of $SL(2,R)$ flat connections can be mapped to solutions of the
Einstein equations with a cosmological constant term,
\be\label{eom}
R_{\mu\nu}=- \,\frac{2}{\ell^2}\,g_{\mu\nu}
\ee
and gauge transformations convert into local frame rotations and
diffeomorphisms.
Strictly speaking,
this is true only on the equations of motion, \emph{e.g.} in the first order formalism,
i.e. for the diffeomorphisms of Einstein geometries.

Chern-Simons theory is a topological theory whose natural observables are monodromies,
that is, path ordered exponentials of the connection over closed loops, \emph{i.e.} Wilson loops.
These gauge invariant quantities measure topological information, they are non-trivial
only in the presence of sources (linkings)  and, on the gravity side,
correspond to horizon lengths (which substitute areas in $3d$).
In classical gravity there is also a richer set of observables:
geodesic lengths between arbitrary points.
It is a natural idea to identify them in some way with {\it open} Wilson lines
in Chern-Simons theory.

As argued in \cite{Wit},
gauge invariance of Chern-Simons theory coincides with diffeomorphism invariance of gravity
modulo Lorentz rotations.
Thus, the geodesic length between two points is related to Wilson average between these points
modulo rotations at the ends, what is exactly the gauge non-invariance of open
Wilson line.
The situation resembles that in knot theory:
Wilson averages in Chern-Simons theory provide invariants of the {\it framed} knots,
knot invariants themselves arise in a particular gauge or for a particular choice of framing
(called topological and defined as a specific property
$H_{_\Box}-1 = O(\hbar^2)$
of the average).
Likewise in the case of 3d gravity the open Wilson lines describe a characteristic
of {\it framed} geodesics, and it turns into geodesic length for a particular choice of gauge.
The need for such freedom is clear from a look at a conical singularity.
In this case, the Wilson loop describes a non-trivial monodromy and its logarithm
is non-vanishing.
At the same time, the geodesic length switches between two branches and vanishes
again when one end approaches the other after walking around the singularity.
The above mentioned gauge freedom in identification of the Wilson average and
the hyperbolic cosine of the geodesic length is exactly what is needed to describe this branch switch.

Quantization of Chern-Simons theories is well understood \cite{WitJones}.
The quantum theory studies expectation values of the Wilson loop operators,
which yield topological invariants of knots and links~\cite{knotpols}.
In this story, the quantum Hilbert space of the Chern-Simons theory
is a space of conformal blocks of certain CFT's, which realize appropriate
representations of the braid groups.
The suggestion of \cite{Wit} was to identify quantum gravity in $3d$
with quantum Chern-Simons theory, and this continues to be
a tantalizing source for various speculations
about the structure of quantum gravity \cite{Carlip}.

An obvious generalization from $SL(2)$ to $SL(N)$ on the Chern-Simons side
can be interpreted as lifting from gravity to higher spin theory,
and, indeed, there is a lot of evidence that analytic continuation in $N$
(i.e. switching to the variable $A=q^N$ in knot polynomials,
which turn into "special" polynomials \cite{MMS} in the classical limit of $q=1$ and $A$ fixed)
has a lot to do with the $3d$ version of higher spin theory \cite{CSgN}.
Even more is expected at the quantum ($q\neq 1$) level.
An enhanced gauge invariance of these theories no longer reduces to diffeomorphism invariance
and this provides non-trivial equivalencies between higher spin theories in
different geometric backgrounds (which are not equivalent from the point of view
of gravity {\it per se}), see \cite{Go} for simple, yet impressive examples.

An additional information on the subject is provided by the AdS/CFT correspondence,
where some gravi\-ta\-tional/Chern-Simons observables should also possess a CFT interpretation.
In the $3d$ case, these are provided by classical (large-$c$) limits of various
conformal blocks and partition functions.
The most spectacular examples so far are interpretations of the geodesic lengths
between points at the AdS boundary (which, according to \cite{JdB,ACI}, are open Wilson loops)
as entanglement entropies in the boundary CFT \cite{Ryu} and as the heavy-light limits
of 4-point classical conformal blocks \cite{Hij}.
From this perspective, the quantization of gravity/high spin theory should lift the AdS/CFT
correspondence to the standard relation between Chern-Simons and Wess-Zumino-Witten conformal theory
\cite{WitJones,Fock}.
Presumably, at this level the controversial open Wilson lines of classical theory
will be promoted to well defined link polynomials in sophisticated representations,
what will add value to further investigation of colored knot/link polynomials.

\bigskip

In the remaining part of this section we discuss what is actually known about the first horizontal arrow in (\ref{GAAG}), namely, about the following maps:

\begin{itemize}
 \item  classical ${\rm AdS}_3$ geometries obtained from flat connections of $3d$ Chern-Simons theories;

\item geodesic lengths from Wilson line integrals, in the cases of flat connections related with the constant ones by a Cartan gauge transformation, when the $P$-exponential coincides with the ordinary one;

\item the bulk geodesic length in the near-boundary limit for the needs of entanglement entropy of $2d$ CFT (Ryu-Takayanagi conjecture \cite{Ryu})
\end{itemize}

\subsection{Metrics from connections}

Let us start from reviewing a class of classical geometries which asymptote to the pure ${\rm AdS}_3$ solution at the conformal boundary of the space-time. We will then remind how these geometries can be obtained from flat connections of the $SL(2,R)\times SL(2,R)$ Chern-Simons theory.

\subsubsection{Metric on ${\rm AdS}_3$}

There exist several conventional parameterizations of anti-de-Sitter space. One common set of coordinates is called global, as it covers the entire space. In the global coordinate patch, we will be interested in the class of metrics with the ${\rm AdS}_3$ boundary conditions (a prescribed decay law sometimes called Brown-Henneaux boundary conditions, \cite{BH}):
\be
\label{metricbc}
\frac{d s^2}{\ell^2}\to d \rho^2 + {\rm e }^{2\rho}\left(-dt^2 + d\phi^2\right) + O(\rho^0)\,, \qquad \rho\to\infty\,,
\ee
where $\rho\to \infty$ is the conformal boundary and $t$ and $\phi$ are cylindrical coordinates $\phi\in [0,2\pi]$.

The simplest representative of this class of metrics is that of pure ${\rm AdS}_3$:
\be
\label{metricAdSg}
\frac{d s^2}{\ell^2} = d\rho^2 - 4\cosh^2\!\rho\,d t^2 + 4\sinh^2\!\rho\, d \phi^2\,.
\ee
It is sometimes convenient to map this patch to the interior of a cylinder via
\be
\frac12\, {\rm e}^{\rho}=\frac{1+\sin\theta}{\cos\theta}\,, \qquad \frac12\,{\rm e}^{-\rho}=\frac{1-\sin\theta}{\cos\theta}\,,
\ee
with $\theta\in[0,\pi/2]$. In terms of $\theta$ the empty AdS metric becomes
\be
\label{metricAdSg2}
\frac{d s^2}{\ell^2} = \frac{1}{\cos^2\theta}\left(d\theta^2 -  4d t^2 + 4\sin^2\theta d\phi^2\right).
\ee
We will also use the Euclidean version of this metric obtained via Wick rotation $t\to i\tau$.

It is sometimes more convenient to use the Poincar\'e patch, which covers a half of the global AdS. In a common choice of the Poincar\'e coordinates
\be
\label{PoincareMetric}
\frac{d s^2}{\ell^2}=\frac{du^2-dt^2+dx^2}{u^2}\,,
\ee
where $u\to 0^+$ corresponds to the region near conformal boundary. In the Euclidean case one can use a complex coordinate parameterization:
\be
\label{PoincareMetricE}
\frac{d s^2}{\ell^2}=\frac{du^2+dzd\bar{z}}{u^2}\,, \qquad z=x+i\tau\,.
\ee

Other asymptotically AdS metrics~(\ref{metricbc}) can be obtained by coordinate rescaling.  As an example, consider the following transformation
\be
\label{diffeo1}
t\to \sqrt{M} t\,, \qquad \phi \to \sqrt{M}\phi\,, \qquad \rho\to \rho-\frac12\,\log M\,,
\ee
with a real positive parameter $M$. This takes~(\ref{metricAdSg}) to
\be
\label{metric1}
\frac{d s^2}{\ell^2} = d\rho^2 - \left(e^{\rho}+M\,e^{-\rho}\right)^2d t^2 + \left(e^{\rho}-M\,e^{-\rho}\right)^2d \phi^2\,.
\ee
This metric also satisfies the Einstein equations~(\ref{eom}) everywhere, except at $\rho_h=1/2\log M$. Indeed at this radius the $\phi$-cycle vanishes and since $\phi$ is $2\pi$-periodic, this metric has a deficit angle $\delta=2\pi(1-\sqrt{M})$. One can understand this deficit angle as created by a source placed at $\rho_h=1/2\log M$. For $M<0$ it the time cycle that becomes contractible and the metric becomes that of a black hole, whose mass is determined by the parameter $M$.

\subsubsection{Connections}

Einstein metrics~(\ref{metric1}) can be related to flat connections of the $SL(2,R)\times SL(2,R)$ \cite{Wit}, or $SL(N,R)\times SL(N,R)$ Chern-Simons theory~\cite{CSgN}. One identifies the dreibein $e_\mu$ and the spin connection $\omega_\mu$ with a linear combinations of the corresponding pair of $SL(N,R)$ connections $A$ and $\bar{A}$:
\be
\label{metric4A}
\frac{e}{\ell}=\frac12\left(A-\bar{A}\right)\,, \qquad \omega = \frac12\left(A+\bar{A}\right)
\ee
The metric and the higher spin fields are then obtained as traces of symmetrized products of $e_\mu$, \emph{e.g.}
\be
g_{\mu\nu}=\frac{1}{{\rm tr}\,T_0^2}\, {\rm tr} \left(e_\mu e_\nu\right), \qquad \phi_{\mu\nu\rho}=\, {\rm tr} \left(e_{(\mu} e_\nu e_{\rho)}\right), \qquad etc.
\ee
Here $T_0$ and $T_{\pm}$ are the generators of $SL(2,R)$ in $SL(N,R)$. It is important how the $SL(2,R)$ are embedded in the $SL(N,R)$ and in what follows the principal embedding will be assumed. We will come back with the details and importance of the embedding in latter sections.

Thus, flat connections are mapped to solutions of the Einstein equations. The gauge transformation acting on the connections are consequently mapped to diffeomorphisms of the metric. Specifically, the infinitesimal transformations
\be
\delta A = d \Lambda + \big[A,\Lambda\big]\,,\qquad \delta \bar{A} = d\bar{\Lambda} + \big[\bar{A},\bar{\Lambda}\big]
\ee
have the following action on the frame fields
\be
\label{SL2diffeos}
\delta e_{\mu}= e_\nu{\xi^\nu}_{;\mu} + \frac12\,\left[e_\mu,\Lambda+\bar{\Lambda}\right]\,, \qquad \xi^\mu =\frac\ell2\,{e^{\mu}}_a\left(\Lambda^a-\bar{\Lambda}^a\right)\,.
\ee

The class of flat connections relevant for~(\ref{metric1}) can be realized as the Cartan gauge transformation with $b=\exp(\rho T_0)$ of the constant connections so that the uppercase flat connections are
\be
\label{transform}
A = b^{-1}(a+\partial)b\,,\qquad \bar{A}=b(\bar a + \partial)b^{-1}
\ee
and the lowercase connections are
\be\label{tria}
a= \left(-T_- +MT_+ + \sum\limits_{s=3}^{N}M_s T^{(s)}_{(s-1)}\right)\left(d\phi+dt\right)\,, \qquad \bar{a}=\left(T_+ - \bar{M}T_-+ \sum\limits_{s=3}^{N}\bar M_s T^{(s)}_{-(s-1)}\right)\left(d\phi-dt\right),
\ee
where $T^{(s)}_{\pm(s-1)}$ are the lowest (highest) weight generators of $SL(N,R)$ classified by the spin $3\leq s\leq N$, and $M_s$ are the corresponding physical charges, classifying the background solution. The $SL(2,R)$ charges $M$ and $\bar{M}$ will be related, in particular, to the Hamiltonian of the gravity background.

In the $SL(2)$ case, for $M=\bar{M}$, adopting the choice of the generators
\be
\label{conventions2}
T_0=\left(
\begin{array}{cc}
1/2 & 0 \\
0 & -1/2
\end{array}\right), \qquad T_-=\left(
\begin{array}{cc}
0 & 0 \\
1 & 0
\end{array}\right), \qquad T_+=\left(
\begin{array}{cc}
0 & 1 \\
0 & 0
\end{array}\right),
\ee
one arrives at the following form of the uppercase connections
\be
\label{ASL2}
A=\left(
\begin{array}{cc}
0 & {\rm e}^{-\rho}M \\
-{\rm e}^{\rho} & 0
\end{array}
\right)dw
+
\left(
\begin{array}{cc}
1/2 & 0 \\
0 & -1/2
\end{array}
\right)d\rho\,,
\ee
\be
\label{AbSL2}
\bar{A}=\left(
\begin{array}{cc}
0 & {\rm e}^{\rho} \\
-{\rm e}^{-\rho}M & 0
\end{array}
\right)d\bar{w}
+
\left(
\begin{array}{cc}
-1/2 & 0 \\
0 & 1/2
\end{array}
\right)d\rho\,,\nn
\ee
where we have introduced complex coordinates
\be
w=\phi +i\tau\,, \qquad \bar{w}=\phi-i\tau\,.
\ee
Using~(\ref{metric4A}) one recovers the Euclidean version of metric~(\ref{metric1}). The Minkowski version is obtained through the Wick rotation $i\tau\to t$.

In the Poincar\'e coordinates the gauge transformation~(\ref{transform}) is generated by
\be
b=u^{-T_0}.
\ee
However, to describe the source in these coordinates the lowercase connections $a$ and $\bar{a}$ must be coordinate dependent. The coordinate dependence enters through the charges $M(z)$, $\bar{M}(\bar{z})$. In the case with no sources $M=\bar M=0$ the gauge connections take the form
\be
\label{APoincareAdS}
A=\left(
\begin{array}{cc}
0 & 0 \\
-1/u & 0
\end{array}
\right)dz
+
\left(
\begin{array}{cc}
-1/2u & 0 \\
0 & 1/2u
\end{array}
\right)du\,,
\ee
\be
\label{AbPoincareAdS}
\bar{A}=\left(
\begin{array}{cc}
0 & 1/u \\
0 & 0
\end{array}
\right)d\bar{z}
+
\left(
\begin{array}{cc}
1/2u & 0 \\
0 & -1/2u
\end{array}
\right)du\,.
\ee
Metric~(\ref{PoincareMetric}) now easily follows from~(\ref{metric4A}).

\subsubsection{Non-constant connections}

The constant lowercase connections (\ref{tria}) are not the most general connections corresponding to asymptotically AdS geometries. In fact, the metric still satisfies the Einstein equations~(\ref{eom}) up to a relevant source term, if one lets the coefficients $M$, $M_s$ be holomorphic functions of $w$ and $\bar M$, $\bar M_s$ of $\bar w$. In the $SL(2)$ case, the connections look like
\be
a = \left(-T_-+J(w)T_+\right)dw \qquad \bar{a}=\left(T_+-\bar{J}(\bar{w})T_-\right)d\bar{w}\,,
\ee
The property $\bar{\partial} J =\partial \bar{J}=0$ is consistent with the flatness condition. The metric reads
\be
{ds^2\over l^2}=d\rho^2 + ({\rm e}^{2\rho}+J+\bar{J}+J\bar{J}{\rm e}^{-2\rho})d\tau^2 + ({\rm e}^{2\rho}-J-\bar{J}+J\bar{J}{\rm e}^{-2\rho})d\phi^2 -2i(J-\bar{J})d\tau\,d\phi\,,
\ee
where the real and imaginary part of $J$ are related to the black hole mass and angular momentum respectively.

The global and Poincar\'e coordinates are related through $z\sim {\rm e}^{\pm iw}$. Thus the sources in the global coordinates, which appeared in the center of the ${\rm AdS}_3$ bulk, are mapped to the sources at $z=0$ (equivalently, at $z=\infty$). As a result, to introduce a source (conical defect) in Poincar\'e coordinates one has to use a coordinate-dependent connection $a$. Then, the metric in the global coordinates is
\be
{ds^2\over l^2} = \frac{du^2}{u^2} + \left(\frac{1}{u}-uJ(z)\right)\left(\frac{1}{u}-u\bar{J}(\bar{z})\right)dx^2 +\left(\frac{1}{u}+uJ(z)\right)\left(\frac{1}{u}+u\bar{J}(\bar{z})\right)dy^2\,.
\ee
For example, taking  $J=\frac{M}{z^2}$, $\bar J=\frac{M}{\bar z^2}$
produces the metric \cite{Hij}
\be
{ds^2\over l^2} = \frac{1}{u^2}\left(d u^2 + \left(1-\frac{Mu^2}{r^2}\right)^2dr^2 + \left(1+\frac{Mu^2}{r^2}\right)^2r^2d\eta^2\right),
\ee
where $r$ and $\eta$ are the polar coordinates on the $z$-plane.

\subsection{Geodesic lengths and Wilson lines}

The connection between Chern-Simons theories and three-dimensional gravity can be further elaborated. In~\cite{JdB,ACI} it was suggested that such natural objects in Chern-Simons theory as Wilson lines (loops) should be interpreted as geodesic distances. Specifically the following relation (valid in the classical limit) was substantiated
\be
\label{WLvsGeodesic}
2\cosh\left({-\sqrt{2c_2(R)}\,\Delta\! s(x_1,x_2)\over l}\right)\ =\ {\rm Tr}_{R}\,{\bf P}\exp\left(-\int_{x_2}^{x_1} A\right) {\bf P}\exp\left(-\int_{x_1}^{x_2} \bar{A}\right) \,.
\ee
Here the geodesic proper distance between points $x_1$ and $x_2$ for a massive probe is compared with a ``composite" Wilson line along a path connecting $x_1$ and $x_2$. The probe's mass is controlled by the quadratic Casimir $c_2$ of the representation $R$, in which the trace is computed, and the contour for holomorphic components $A$ and $\bar{A}$ is traversed in opposite directions.\footnote{A somewhat independent motivation for this Keldysh-like choice of the contour came recently from the work~\cite{Nair}, where it was demonstrated how Chern-Simons fields appear from a thermofield description of a quantum system.}

Two considerations immediately arise when looking at~(\ref{WLvsGeodesic}). First, the geodesic path is almost uniquely specified by the geometry, while the Wilson lines of flat connections are path independent. Second, Wilson lines are gauge dependent objects, they transform non-locally under the gauge group action. One may wonder how this is resolved on the gravity side.

As far as the choice of path for the geodesics is concerned it is in fact a gauge freedom. In~\cite{ACI} it was explained how the proper geodesic path can appear from a convenient choice of the gauge. Meanwhile the problem of gauge  dependence is controlled by the specific configuration of the Wilson line, in this case the composite Wilson line above~\cite{JdB}. This particular choice is invariant under the diagonal $SL(2,R)\times SL(2,R)$ transformations $\Lambda=\bar{\Lambda}$, which from the point of view of~(\ref{SL2diffeos}) are Lorentz frame rotations, leaving the metric invariant. On the other hand the non-diagonal transformations that do transform the composite Wilson line correspond to those diffeomorphisms that change the metric:
\be
\delta g_{\mu\nu} = \xi_{\mu;\nu}+\xi_{\nu;\mu}
\ee

To show how things work, let us review some examples of the correspondence~(\ref{WLvsGeodesic}).

First, in the purely topological case, when there is no matter, consequently $c_2(R)=0$, the right hand side of~(\ref{WLvsGeodesic}) is one. Accordingly, the Wilson operator in the trivial representation is just a number, independent of $x_1$ or $x_2$.

\subsubsection{Pure AdS}

\paragraph{Poincar\'e coordinates.} In the Poincar\'e AdS~(\ref{PoincareMetric}) the geodesics are simply half-circles anchored at the boundary of the AdS:
\be
u(x)=\sqrt{R^2-(x-x_0)^2}\,,
\ee
where the radius of the circle can be expressed in terms of the distance $\Delta x$ between the endpoints of the half-circle at the boundary, $R=\Delta x/2$. Thus, the general geodesic distance is the length of the arc connecting two points in the bulk. Below we parameterize the geodesic distance by the hyperbolic parameter
\be
\label{gdistance}
\Delta s(X,Y)\equiv\ell\, \log \left(\zeta+\sqrt{\zeta^2-1}\right),\ \ \ \ \ \hbox{i.e.}\ \ \ \ \ \cosh\left({\Delta s(X,Y)\over l}\right)=\zeta
\ee
For two generic points $X=(\epsilon_1,x_1,0)$ and $Y=(\epsilon_2,x_2,0)$ the length would be given by
\be
\label{zPoincare}
\zeta= \frac{1}{2}\left( \frac{\epsilon_1}{\epsilon_2} + \frac{\epsilon_2}{\epsilon_1} + \frac{(x_1-x_2)^2}{\epsilon_1\epsilon_2}\right)\,.
\ee

Let us compare this result for the geodesic distance with the computation of a relevant Wilson line. Since the Wilson line is path independent we select the path to consist of two segments: from $X$ to $Z=(\epsilon_1,x_2,0)$ with constant $z=\epsilon_1$ and from $Z$ to $Y$ with constant $x=x_2$,
\begin{multline}
{\bf P}\exp\left(- \int_{X}^Y \bar{A}\right) = {\bf P}\exp\left(-\int_Z^Y \bar{A}_u\,du \right) {\bf P}\exp \left(-\int_X^Z \bar{A}_x\,dx\right) \\ =
\left(
\begin{array}{cc}
\sqrt{{\epsilon_1}/{\epsilon_2}} & 0 \\
0 & \sqrt{{\epsilon_2}/{\epsilon_1}}
\end{array}
\right)
\left(
\begin{array}{cc}
1 & -\delta x/\epsilon_1 \\
0 & 1
\end{array}
\right)
 =
\left(
\begin{array}{cc}
\sqrt{{\epsilon_1}/{\epsilon_2}} & -\delta x/\sqrt{{\epsilon_1\epsilon_2}} \\
0 & \sqrt{{\epsilon_2}/{\epsilon_1}}
\end{array}
\right),
\end{multline}
where $\delta x=x_2-x_1$. Similarly,
\be
{\bf P}\exp\left(- \int_{Y}^X A\right) =
\left(
\begin{array}{cc}
1 & 0 \\
-\delta x/\epsilon_1 & 1
\end{array}
\right)
\left(
\begin{array}{cc}
\sqrt{{\epsilon_1}/{\epsilon_2}} & 0 \\
0 & \sqrt{{\epsilon_2}/{\epsilon_1}}
\end{array}
\right)= \left(
\begin{array}{cc}
\sqrt{{\epsilon_1}/{\epsilon_2}} & 0 \\
-\delta x/\sqrt{{\epsilon_1\epsilon_2}} & \sqrt{{\epsilon_2}/{\epsilon_1}}
\end{array}
\right).
\ee
The product of the holomorphic and antiholomorphic parts yields
\be
\left(
\begin{array}{cc}
{{\epsilon_1}/{\epsilon_2}} & -\delta x/\epsilon_2 \\
-\delta x/\epsilon_2 & \epsilon_2/\epsilon_1+ \delta x^2/(\epsilon_1\epsilon_2)
\end{array}
\right)
\ee
Here we used the generators in the fundamental representation of $SL(2)$. After taking the trace, one gets
\be
W_\Box(X,Y)=  \frac{\epsilon_1^2 + \epsilon_2^2 + (x_2-x_1)^2}{\epsilon_1\epsilon_2}\,,
\ee
Therefore one finds the following relation between the two results:
\be
\label{WLvsGeodesic2}
W_\Box(X,Y)= 2\cosh\frac{\Delta s(X,Y)}{\ell}.
\ee
In the AdS/CFT applications, such as Ryu-Takayanagi formula, one need to take $\epsilon_1\sim\epsilon_2\ll 1$.

\paragraph{Global coordinates.} The same exercise can be done in the global coordinates. In the pure AdS space, specified by metric~(\ref{metricAdSg}), an equal time geodesic can be presented by the curve
\be
\phi-\phi_0 = \frac12\,{\rm arctan}\,\frac{C\cosh\rho}{\sqrt{\sinh^2\rho-C^2}}\,,
\ee
where $C$ is the integration constant. The geodesics distance between two arbitrary points is again provided by~(\ref{gdistance}) with
\be
\label{ds2}
\zeta = \cosh\rho_1\cosh\rho_2\cos2(t_1-t_2)-\sinh\rho_1\sinh\rho_2\cos2(\phi_1-\phi_2)\,.
\ee
Choosing $X=(\log\epsilon_1,0,0)$ and $Y=(\log\epsilon_2,0,\phi)$, one gets
\be
\zeta
=\frac{(\epsilon_1^2+\epsilon_2^2)\cos^2\!\phi+(1+\epsilon_1^2\epsilon_2^2)\sin^2\!\phi}{2\epsilon_1\epsilon_2}\,.
\ee

Now we proceed with the Wilson loop computation using connections~(\ref{ASL2}) and~(\ref{AbSL2}). The anti-holomorphic part of the Wilson line connecting points $X$ and $Y$ reads
\begin{multline}
{\bf P}\exp\left(- \int_{X}^Y \bar{A}\right) = {\bf P}\exp\left(-\int_Z^Y \bar{A}_\rho\,d\rho\right) {\bf P}\exp \left(-\int_X^Z \bar{A}_\phi\,d\phi\right)=
\\ =
\left(
\begin{array}{cc}
\sqrt{{\epsilon_2}/{\epsilon_1}} & 0 \\
0 & \sqrt{{\epsilon_1}/{\epsilon_2}}
\end{array}
\right)
\left(
\begin{array}{cc}
\cos\phi & -\epsilon_1\sin\phi \\
\ds \frac{1}{\epsilon_1}\sin\phi & \cos\phi
\end{array}
\right)
=
\left(
\begin{array}{cc}
\sqrt{{\epsilon_2}/{\epsilon_1}}\cos\phi & -\sqrt{{\epsilon_1\epsilon_2}}\sin\phi \\
\sqrt{{1}/{\epsilon_1\epsilon_2}}\sin\phi & \sqrt{{\epsilon_1}/{\epsilon_2}}\cos\phi
\end{array}
\right),
\end{multline}
where we first went from $X$ to $Z=(\log\epsilon_1,0,\phi)$ at constant $\rho=\log\epsilon_1$, and then, from $Z$ to $Y$ at constant $\phi$. For the holomorphic part one gets
\begin{multline}
{\bf P}\exp\left(- \int_{Y}^X {A}\right) =
\left(
\begin{array}{cc}
\cos\phi & \ds \frac{1}{\epsilon_1}\sin\phi \\
\ds -\epsilon_1\sin\phi & \cos\phi
\end{array}
\right)
\left(
\begin{array}{cc}
\sqrt{{\epsilon_2}/{\epsilon_1}} & 0 \\
0 & \sqrt{{\epsilon_1}/{\epsilon_2}}
\end{array}
\right)
=
\left(
\begin{array}{cc}
\sqrt{{\epsilon_2}/{\epsilon_1}}\cos\phi & \sqrt{{1}/{\epsilon_1\epsilon_2}}\sin\phi \\
-\sqrt{{\epsilon_1\epsilon_2}}\sin\phi & \sqrt{{\epsilon_1}/{\epsilon_2}}\cos\phi
\end{array}
\right).
\end{multline}
In the fundamental representation the trace of the product of the holomorphic and anti-holomorphic contributions yields
\be
W_\Box(X,Y) = \frac{(\epsilon_1^2+\epsilon_2^2)\cos^2\!\phi+(1+\epsilon_1^2\epsilon_2^2)\sin^2\!\phi}{\epsilon_1\epsilon_2}
\ee
Again, we see that relation~(\ref{WLvsGeodesic2}) holds.

\subsubsection{Connections with conical singularity}

So far we considered a topologically trivial case of the pure AdS. Our next example is the metric with a conical singularity. We use connection (\ref{ASL2}) which gives the Euclidean version of~(\ref{metricAdSg2}). This is a metric with a conical singularity at $\rho=\log \sqrt{M}$ with a deficit angle $2\pi(1-\sqrt{M})$.
We compute the Wilson loop of the connection $A$ around the conical singularity
\be
{\bf P}\exp\left(-\oint A_\phi\ d\phi\right)=  \left(
\begin{array}{cc}
\cos\left(2\sqrt{M}\pi\right) & -\sqrt{M}{{\rm e}^{-\rho}}\,\sin\left(2\sqrt{M}\pi\right) \\
\ds \frac{{\rm e}^{\rho}}{\sqrt{M}}\,\sin\left(2\sqrt{M}\pi\right) & \cos\left(2\sqrt{M}\pi\right)
\end{array}
\right)
\ee
This monodromy is non-trivial unless $\sqrt{M}$ is a half-integer number.

For the Wilson line with $\rho_1=\rho_2$, $\Delta\phi=\phi_2-\phi_1$ the result simply generalizes to
\be
W(\phi_1,\phi_2)={\bf P}\exp\left(-\oint A_\phi\ d\phi\right)=  \left(
\begin{array}{cc}
\cos\left(\sqrt{M}\Delta\phi\right) & -\sqrt{M}{{\rm e}^{-\rho}}\,\sin\left(\sqrt{M}\Delta\phi\right) \\
\ds \frac{{\rm e}^{\rho}}{\sqrt{M}}\,\sin\left(\sqrt{M}\Delta\phi\right) & \cos\left(\sqrt{M}\Delta\phi\right)
\end{array}
\right).
\ee
Since the space is not simply connected there are two branches of the Wilson line, depending on the direction (clockwise/counterclockwise) the path encircles the singularity. The corresponding Wilson lines are related via a transformation $\phi\to \phi-2\pi$, which can be realized as a $SO(2)$ gauge transformation via a left and/or right multiplication, \emph{e.g.}
\be
W_2(\phi_1,\phi_2)=\left(
\begin{array}{cc}
\cos\left(2\sqrt{M}\pi\right) & \sqrt{M}{{\rm e}^{-\rho}}\,\sin\left(2\sqrt{M}\pi\right) \\
\ds -\frac{{\rm e}^{\rho}}{\sqrt{M}}\,\sin\left(2\sqrt{M}\pi\right) & \cos\left(2\sqrt{M}\pi\right).
\end{array}
\right) W_1(\phi_1,\phi_2)\,,
\ee
where the transformation is understood as acting on the endpoint $\phi_2$ via a shift by $2\pi$.

Similarly, for the geodesics, there are two distances to compute. Using~(\ref{ds2}), the distance between two points with the same $\rho$ and $\tau$ in empty AdS, is given by
\be
\zeta=\cosh^2\!\rho-\sinh^2\!\rho\,\cos2\Delta\phi\,,
\ee
which is invariant under the transformation $\phi\to \phi-2\pi$. This is not true if there is a conical singularity, and there are two geodesic distances, which distinguish which direction the singularity is bypassed:
\be
\zeta_1(\phi_1,\phi_2)=\cosh^2\!\rho-\sinh^2\!\rho\,\cos\left(2\sqrt{M}\Delta\phi\right)\\
\zeta_2(\phi_1,\phi_2)=\cosh^2\!\rho-\sinh^2\!\rho\,\cos\left(2\sqrt{M}\left(\Delta\phi-2\pi\right)\right)=\zeta_1(\phi_1,\phi_2-2\pi).
\ee
Clearly the same shift applies if $\rho_1\neq \rho_2$.

\subsection{Living on the AdS boundary}

So far we discussed relations between three dimensional gravity and Chern-Simons theory, i.e. those of type (\ref{geodlength=mamo}). For these relations one should not put the endpoints of the geodesics (or Wilson line) at the boundary of the ${\rm AdS}_3$ space. However, these relations has nothing to do with the AdS/CFT correspondence.

Another point of interest are relations of the other, (\ref{adscft}) type. Those are due to the ${\rm AdS}_3/{\rm CFT}_2$ correspondence: the endpoints of the Wilson lines should be anchored at the boundary, $\rho=-\log\epsilon$, with $\epsilon\to 0$. The initial point is taken to be at the origin, while $w$ stays for the final point. We will be further interested in that kind of Wilson lines.

Following \cite{HKP}, let us rewrite the uppercase $A$ and $\bar{A}$ in terms of the $SL(N)$ group elements
\be
A=L(x)dL^{-1}(x)\,, \qquad \bar{A} = R^{-1}(x)dR(x)
\ee
which in terms of the coordinates of metric~(\ref{metric1}) are
\be
L={\rm e}^{-\rho T_0}{\rm e}^{a_ww+a_{\bar{w}}\bar{w}}\,, \qquad R={\rm e}^{\bar{a}_ww+\bar{a}_{\bar{w}}\bar{w}}{\rm e}^{-\rho T_0}
\ee
The combined Wilson line is
\be
W={\bf P}\exp\left(-\int_{\bar C} A\right)\, {\bf P}\exp\left(-\int_C {\bar A}\right)=L(x_i)L^{-1}(x_f)R^{-1}(x_f)R(x_i)\,,
\ee
where $C$ is a contour going from $x_i$ to $x_f$, ${\bar C}$ is the reversed contour.

Thus, we need to evaluate the traces of $W$ in arbitrary representations $R$
\be\label{W}
W_R(C)={\rm Tr}_R W = {\rm Tr}_R \left({\rm e}^{-\log\epsilon T_0}{\rm e}^{-a_ww-a_{\bar{w}}\bar{w}}{\rm e}^{2\log\epsilon T_0}{\rm e}^{\bar{a}_ww+\bar{a}_{\bar{w}}\bar{w}}{\rm e}^{\log\epsilon T_0}\right)
\ee
at $\epsilon\to 0$. To this end, we introduce projectors $P_\pm$ onto the highest/lowest vector of the representation $R$:
\be
P_\pm =|\pm{\rm hw}\,\rangle\langle\,\pm{\rm hw}\,|=\lim\limits_{\epsilon\to 0}\epsilon^{ 2h_R}{\rm e}^{\mp2\log\epsilon T_0}\,, \qquad T_0|\pm{\rm hw}\,\rangle= \pm h_R|\pm{\rm hw}\,\rangle
\ee
and note that
\be
\label{Wfactorized}
W_R(C)={\rm e}^{-4h_R}{\rm Tr}_R \left(P_+{\rm e}^{-a_ww-a_{\bar{w}}\bar{w}}P_-{\rm e}^{\bar{a}_ww+\bar{a}_{\bar{w}}\bar{w}}\right)={\rm e}^{-4h_R}\langle\,{\rm hw}\,|{\rm e}^{-a_ww-a_{\bar{w}}\bar{w}}|-{\rm hw}\,\rangle\langle\,-{\rm hw}\,|{\rm e}^{\bar{a}_ww+\bar{a}_{\bar{w}}\bar{w}}|{\rm hw}\,\rangle
\ee
as $\epsilon\to 0$. Thus of the main interest will be the matrix elements of ${\rm e}^\Lambda={\rm e}^{-a_ww-a_{\bar{w}}\bar{w}}$ between the highest and the lowest states.

In fact, in equation~(\ref{Wfactorized}) the result comes in a form with the contributions of the left and right sectors have factorized, we will be finally interested only in the matrix element of ${\rm e}^\Lambda={\rm e}^{-\bar a_w\bar w}$ with $\bar a_w$ as in (\ref{tria}).

\subsection{Summary \label{sum2}}

As we briefly surveyed in this section, the common lore in the field amounts to the following:

$\bullet$ Really interesting observables to consider
in $3d$ classical higher-spin/Chern-Simons theory are
open Wilson lines, i.e. the $P$-exponentials of flat connections.

$\bullet$ It is already interesting to consider the flat connections (\ref{transform}) induced by {\it constant} connections, (\ref{tria}), then the $P$-exponentials turn into ordinary matrix exponents.

$\bullet$ Of primary interest are their matrix elements between the highest and/or lowest
 vectors of various representations $R$, then it is enough to study fundamental representations,
 because {\it such} matrix elements factorize according to (\ref{clafactor}).

\noindent

Common justification for all these points comes from the AdS/CFT studies,
where the AdS boundary plays a special role, and it is possible to restrict gauge
invariance to transformations with the same asymptotics, i.e. constant at the boundary.
Then, for the Wilson lines with the ends at the boundary
invariant are not only traces, but also their eigenvalues (modulo permutations).
Restriction to highest/lowest states is then also natural, because the Wilson lines can diverge
and such matrix elements provide the main contribution in the near-boundary limit~\cite{HKP}.
This justifies the following task:

$\bullet$ It deserves looking at  families of exponentiated constant connections,
which are fully determined by their eigenvalues,
and evaluate their corner matrix elements in all fundamental representations.

As we shall see in the next section, matrix elements distinguished by the above properties
are also distinguished  by their close relation to  the standard integrability theory:
as often happens, the quantity interesting for physical needs is exactly the one that
is covered by powerful mathematical methods.

%\newpage

\section{$\tau$-functions\label{math}}

\subsection{Generalities}

We now switch to a more formal investigation of the problem.
The task is to find a description of matrix elements of $e^\Lambda$
in arbitrary representation $R$.
In general, representation dependence is a subject of theory of
(non-Abelian) $\tau$-functions \cite{GKLMM} and is attracting an increasing attention
(e.g. in the form of theory of $\hat A$-polynomials in knot theory \cite{Apol}).
It is an unexpectedly hard problem and
still an underdeveloped field.
Fortunately, in the present context there is a number of simplifications,
provided exactly by the items of sec.\ref{sum2}:

$\bullet$ Restriction to constant connections allows one to consider
group elements of ordinary Lie algebras, neither Kac-Moody, nor quantum
extensions are immediately needed.

$\bullet$ Restriction to matrix elements between highest/lowest weight
states reduces through (\ref{clafactor}) the problem to that in the fundamental representations,
what in the $\tau$-function language means restriction to Toda $\tau$-functions.
The simplification is considerable: matrix elements in higher fundamental representations
are just minors of those in the first one.

$\bullet$ The peculiar form (\ref{tria}) of the connection $\Lambda$
implies that for $n$ non-vanishing "times" $w_s$
the study of $n\times n$ matrices can be actually
sufficient to solve the problem for arbitrary $N$.
This is still a  trivial procedure for $n=2$, because with (\ref{tria}) we deal with the
$SL(2)$ subgroup of $SL(N)$, but it is less trivial for $n>2$, since (\ref{tria})
is in no way a special embedding  $SL(n)\subset SL(N)$ (which simply can not exist),
still there should be a "lifting procedure" from $n\times n$ matrices to $N\times N$,
generalizing the one for $n=2$.

\bigskip

Technically, this implies the following strategy.

Conceptually,  $\tau$-function is defined \cite{GKLMM} as a generating function of all matrix
elements
\be
\tau_R(t,\bar t|G) = \sum_{\mu,\nu} <\mu|G|\nu>_R t^{\mu}\bar t^{\nu}
\label{taudef}
\ee
where $G$ is a group element, i.e. the representation dependence
is controlled by the comultiplication rule $\Delta(G)=G\otimes G$.
For quantum groups this means that $G$ and thus $\tau$ is operator valued
\cite{GKLMM,MorVin,Mir},
but within classical higher spin theory we do not need this (difficult) generalization.
According to \cite{GKLMM}, the $\tau$-functions (\ref{taudef}) always satisfy bilinear
Hirota equations,
which are just avatars of the representation multiplication
$R_1\otimes R_2 = \sum_Q C_{R_1R_2}^Q Q$, rewritten in a form of
differential/difference equations in variables $t,\bar t$.
The difficult problem in the theory of $\tau$-function is a proper
choice of the generating function, i.e. parametrization of the states $\mu$ and $\nu$
within representation $R$.

In the simplest case of the $n$-th fundamental representation,
one can generate all of them by the action of the-first-subdiagonal generator $T_+$ and its degrees, which gives a set of commuting "Hamiltonians" $H$,
\be
\tau_{[n]} = <{\rm hw}| e^{H(t)}G e^{\bar H(\bar t)}|{\rm hw}>,\ \ \ \ \ \ \ H(t)=\sum_k t_kT_+^k,\ \ \ \ \ \ \bar H(\bar t)=\sum_k\bar t_kT_-^k
\label{tauHam}
\ee
However, in other representations this does not produce {\it all} the states in $R$,
this violates completeness and jeopardizes the Hirota equations, which relate matrix elements in the fundamental representations at different $n$'s. Hence, in these cases one has \cite{GKLMM} to generalize (\ref{tauHam}) and to consider the generic Gauss decomposition,
\be\label{tau}
\tau = <{\rm hw}| e^{U(t)}G e^{L(\bar t)}|{\rm hw}>
\ee
where $L$ and $U$ are {\it generic} Borel (lower and upper triangular) matrices,
i.e. associate time-variables with non-Abelian, still nilpotent subgroups. One can definitely consider other possibilities to have generating functions of all matrix elements (\ref{taudef}) which correspond to complicated non-linear changes of time variables in (\ref{tau}):
\be\label{skewtau}
\tau_- = <{\rm hw}| e^{U(t)}G e^{U(\bar t)}|-{\rm hw}>\nn\\
\tau_-^* = <-{\rm hw}| e^{L(t)}G e^{L(\bar t)}|{\rm hw}>\nn\\
\tau_= = <-{\rm hw}| e^{L(t)}G e^{U(\bar t)}|-{\rm hw}>
\ee
$|-{\rm hw}>$ denotes here the lowest weight vector and we call these $\tau$-functions skew and double skew respectively. They sometimes still satisfy similar Hirota like bilinear identities, however, the main problem with the skew $\tau$-functions is that all the formulas are hardly continued from fixed to arbitrary $N$, in contrast with the standard $\tau$-functions. It is related to the fact that a continuation of the lowest weight vector for arbitrary $N$ is done much harder than the highest weight vector.

Coming back to our problem, to put it to a $\tau$-function language
one needs to convert $e^\Lambda$ into a Gauss decomposition form.
This is a sophisticated non-linear transformation of time variables,
$(u,w_s)\longrightarrow (t,\bar t)$.
However, since $w_s$ are defined as coefficients in front of the algebra {\it generators},
it can be performed, at least in principle, with the help of the
Campbell-Hausdorff formula, i.e. the transformation is "functorial": it
does not depend on the representation $R$.
Moreover,  for $n=2$ the non-vanishing $w_s$ the generators belong to $SL(2)\subset SL(N)$,
and one can perform the transformation (Gauss decomposition) just for $2\times 2$ matrices:
it will be automatically continued to arbitrary $N$.
This means that all the multiple commutators in the Campbell-Hausdorff formula
are expressed through just three ones: the generators of $SL(2)$.
Unfortunately, this property disappears for $n>2$: multiple commutators of
$n^2-1$ generators $V_i^{(s)}$ are {\it not} expressed through themselves,
and the problem requires a more sophisticated approach.

In the next sections, we provide some more details about these procedures
and demonstrate that the answers for the skew $\tau$-functions indeed possess certain properties similar to those of
the standard $\tau$-functions.
This approach can help to systematize somewhat sporadic observations
in the literature; this paper is, however, just a first step in this direction.

\subsection{$SL(2)$ with $SL(N)$}

We begin by reminding some basic facts from
the elementary representation theory of $SL(N)$ Lie algebras (see also \cite{GKLMM}).
Occasionally, this subsection is almost the same as Appendix A of \cite{Go}.

\bigskip

Representation $[p]$ of principally embedded
$SL(2)$ by $(p+1)\times (p+1)$ matrices looks as follows:

\bigskip

{\footnotesize
\be
E^{(p)} = \left(\begin{array}{ccccccccc}
0 &\sqrt{p} \\
& 0 & \sqrt{2(p-1)}\\
&&&\ldots\\
&&&& 0 & \sqrt{i(p-i+1)}\\
&&&&&0\\
&&&&&&\ldots\\
&&&&&&&0&\sqrt{p} \\
&&&&&&&&0
\end{array}\right)
\nn\\
F^{(p)} = \left(\begin{array}{ccccccccc}
0 \\
\sqrt{p} & 0 \\
& \sqrt{2(p-1)}& 0\\
&&&\ldots\\
&&&& 0 \\
&&&& \!\!\!\!\!\!\!\!\!\!\sqrt{i(p-i+1)} & 0\\
&&&&&&\ldots\\
&&&&&&&0 \\
&&&&&&&\sqrt{p}&0
\end{array}\right)
\nn\\
H^{(p)}  = \frac{1}{2}\left(\begin{array}{ccccccccc}
p \\
& p-2\\
&& p-4 \\
&&&\ldots\\
&&&&&-p
\end{array}\right)
\ee}
Then
\be
[H,E]=E, \ \ [H,F]=-F,\ \ [E,F]=2H\ee

Thus, in the fundamental representation of $SL(r+1)$
\be
F^{(r)} = \sum_{i=1}^r \sqrt{i(r+1-i)}\cdot T_{-\alpha_i}
\ee
In what follows we denote this linear combination in arbitrary
representation by $T_-$:
\be
T_- \equiv \sum_{i=1}^r \sqrt{i(r+1-i)}\cdot T_{-\alpha_i}
\label{defT}
\ee
Together with
\be
T_+ \equiv \sum_{i=1}^r \sqrt{i(r+1-i)}\cdot T_{\alpha_i}
\ee
and
\be
H \equiv \sum_{i=1}^r i(r+1-i)\cdot H_{\alpha_i}
\ee
these generators form an $SL(2)$ subalgebra of $SL(N)$.
Other generators of $SL(N)$ are naturally decomposed into
integer spin representations
w.r.t. this subalgebra:
\be
\begin{array}{c|c|cc}
\text{spin} & \text{dimension} & \text{highest vector} \\  &&\\
1  & 3 & T_+ =\sum_{i=1}^r \sigma_i \cdot T_{\alpha_i}\\ &&\\
2  & 5 & T^{(2)}_+= \sum_{i=1}^{r-1} \sigma_i\sigma_{i+1}\cdot T_{\alpha_i+\alpha_{i+1}} \\         &&\\
3 &  7 &  T^{(3)}_+=\sum_{i=1}^{r-2} \sigma_i\sigma_{i+1}\sigma_{i+2} \cdot
T_{\alpha_i+\alpha_{i+1}+\alpha_{i+2}} \\ &&\\
\ldots    && \\ && \\
r & 2r+1 & T^{(r)}_+ =  T_{\alpha_1+\ldots + \alpha_r}
\end{array}
\ee
Here $\sigma_i = \sqrt{i(r+1-i)}$.
The highest weight vectors $T^{(s)}_+$ are those annihilated by commutating with $T_+$,
they have entries on a given upper subdiagonal.
All other elements of the representation are obtained by repeated
commutation of highest weight vector with $T_-$ (by adjoint action of $T_-$).
The lowest weight vector in the representation is
\be
T^{(s)}_- = \sum_{i=1}^{r+1-s} \left(\prod_{j=1}^{s}
\sigma_{i+j-1} \right)\cdot T_{-\alpha_i-\ldots - \alpha_{i+s-1}}
\ee

\section{Standard $\tau$-function}

Now we are ready to describe in detail what is the standard $\tau$-function, the issue briefly touched in s.1.1.

\subsection{Extract from \cite{GKLMM}}

The first fundamental representation
\be
F=\{ \psi_i = T_-^{i}\psi_0\ | \ i=0..r\}
\ee
is made from the highest weight $\ |\text{hw}_F>=\psi_0\ $ by a single generator
(\ref{defT}).
Note that because of non-trivial coefficients in (\ref{defT}) the states $\psi_i$
are non-trivially normalized:
\be
<\psi_i|\psi_j> \ = \delta_{i,j}\cdot{\prod_{k=1}^i k(r+1-k)}
= \frac{i!r!}{(r-i)!}\cdot \delta_{i,j}
\ee
in particular,
\be
<\psi_r|T_-^r|\psi_0>\ =\ <\psi_r|\psi_r>\  = {(r!)^2}
\ee
\be\label{ren}
<\psi_r |e^{\bar t T_-}|\psi_0> \  = \bar t^r\cdot r!
\ee
and
\be
<\psi_0|e^{tT_+}e^{\bar tT_-}|\psi_0> \
= \sum_{i=0}^r \cdot \frac{(t\bar t)^i}{(i!)^2}\frac{i!r!}{(r-i)!}
= \left(1+t\bar t\right)^r
\label{exatau1}
\ee
where we assumed that $\  <\psi_0|\psi_0>\ \equiv 1$.
The operators $T^{(s)}_\pm$ act in $F$ just as the $s$-th powers of $T^\pm$:
\be
T^{(s)}_\pm = T_\pm^s\Big|_{\text{on}\ F}
\label{Tspowers}
\ee

\bigskip

Higher fundamental representations
\be
F_k = \Lambda^k(F) = \{\psi_{[i_1}\ldots \psi_{i_k]} \ | \ 0\leq i_1 < i_2 < \ldots < i_k\leq r\}
\ee
with $k=1,\ldots,r$
are made from their highest weights $\ |\text{hw}_k>=\psi_{[0}\psi_{1}\ldots \psi_{k]}\ $
by $k$ generators
\be
\Delta_k(T_-^i) = T_-^i\otimes I\otimes \ldots \otimes I +
I\otimes T_-^i \otimes \ldots \otimes I + \ldots + I\otimes I \otimes \ldots \otimes T_-^i,
\ \ \ \ \ \ \ \ \ i=1,\ldots,k
\ee

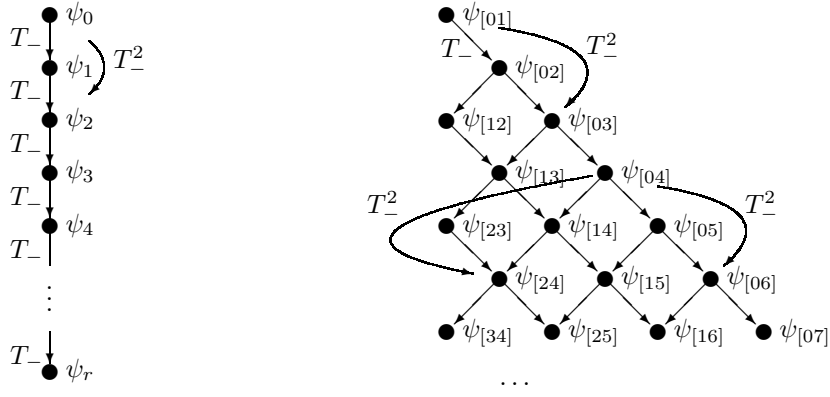
\begin{picture}(300,200)(-150,-180)
\put(-50,0){\circle*{6}}\put(-44,-2){\mbox{$\psi_0$}}
\put(-50,-20){\circle*{6}}\put(-44,-22){\mbox{$\psi_1$}}
\put(-50,-40){\circle*{6}}\put(-44,-42){\mbox{$\psi_2$}}
\put(-50,-60){\circle*{6}}\put(-44,-62){\mbox{$\psi_3$}}
\put(-50,-80){\circle*{6}}\put(-44,-82){\mbox{$\psi_4$}}
\put(-51.5,-112){\mbox{$\vdots$}}
\put(-50,-135){\circle*{6}}\put(-44,-137){\mbox{$\psi_r$}}
\put(-50,0){\vector(0,-1){17}}
\put(-50,-20){\vector(0,-1){17}}
\put(-50,-40){\vector(0,-1){17}}
\put(-50,-60){\vector(0,-1){17}}
\put(-50,-80){\line(0,-1){15}}
\put(-50,-120){\vector(0,-1){12}}
\put(-65,-12){\mbox{$T_-$}}
\put(-65,-32){\mbox{$T_-$}}
\put(-65,-52){\mbox{$T_-$}}
\put(-65,-72){\mbox{$T_-$}}
\put(-65,-92){\mbox{$T_-$}}
\put(-65,-132){\mbox{$T_-$}}
\qbezier(-35,-10)(-24,-20)(-35,-30)\put(-33,-28){\vector(-1,-1){2}}\put(-26,-20){\mbox{$T_-^2$}}
\put(100,0){\circle*{6}}\put(106,-2){\mbox{$\psi_{[01]}$}}
\put(120,-20){\circle*{6}}\put(126,-22){\mbox{$\psi_{[02]}$}}
\put(100,-40){\circle*{6}}\put(106,-42){\mbox{$\psi_{[12]}$}}
\put(140,-40){\circle*{6}}\put(146,-42){\mbox{$\psi_{[03]}$}}
\put(120,-60){\circle*{6}}\put(126,-62){\mbox{$\psi_{[13]}$}}
\put(160,-60){\circle*{6}}\put(166,-62){\mbox{$\psi_{[04]}$}}
\put(100,-80){\circle*{6}}\put(106,-82){\mbox{$\psi_{[23]}$}}
\put(140,-80){\circle*{6}}\put(146,-82){\mbox{$\psi_{[14]}$}}
\put(180,-80){\circle*{6}}\put(186,-82){\mbox{$\psi_{[05]}$}}
\put(120,-100){\circle*{6}}\put(126,-102){\mbox{$\psi_{[24]}$}}
\put(160,-100){\circle*{6}}\put(166,-102){\mbox{$\psi_{[15]}$}}
\put(200,-100){\circle*{6}}\put(206,-102){\mbox{$\psi_{[06]}$}}
\put(100,-120){\circle*{6}}\put(106,-122){\mbox{$\psi_{[34]}$}}
\put(140,-120){\circle*{6}}\put(146,-122){\mbox{$\psi_{[25]}$}}
\put(180,-120){\circle*{6}}\put(186,-122){\mbox{$\psi_{[16]}$}}
\put(220,-120){\circle*{6}}\put(226,-122){\mbox{$\psi_{[07]}$}}
\put(100,0){\vector(1,-1){17}}\put(98,-16){\mbox{$T_-$}}
\put(120,-20){\vector(1,-1){17}}
\put(120,-20){\vector(-1,-1){17}}
\put(100,-40){\vector(1,-1){17}}
\put(140,-40){\vector(1,-1){17}}
\put(140,-40){\vector(-1,-1){17}}
\put(120,-60){\vector(1,-1){17}}
\put(120,-60){\vector(-1,-1){17}}
\put(160,-60){\vector(1,-1){17}}
\put(160,-60){\vector(-1,-1){17}}
\put(100,-80){\vector(1,-1){17}}
\put(140,-80){\vector(1,-1){17}}
\put(140,-80){\vector(-1,-1){17}}
\put(180,-80){\vector(1,-1){17}}
\put(180,-80){\vector(-1,-1){17}}
\put(120,-100){\vector(1,-1){17}}
\put(120,-100){\vector(-1,-1){17}}
\put(160,-100){\vector(1,-1){17}}
\put(160,-100){\vector(-1,-1){17}}
\put(200,-100){\vector(1,-1){17}}
\put(200,-100){\vector(-1,-1){17}}
\put(120,-140){\mbox{$\ldots$}}
\qbezier(120,-5)(170,-13)(145,-35)\put(147,-33){\vector(-1,-1){2}}\put(153,-15){\mbox{$T_-^2$}}
\qbezier(180,-65)(230,-73)(205,-95)\put(207,-93){\vector(-1,-1){2}}\put(213,-75){\mbox{$T_-^2$}}
\qbezier(155,-61)(30,-83)(110,-98)\put(108,-97.5){\vector(4,-1){2}}\put(70,-75){\mbox{$T_-^2$}}
\end{picture}

\noindent
The "depth" of the representation $F_k$ from $\text{hw}_k=\psi_{[0,\ldots, {k-1}]}$
to $-\text{hw}_{-k}=\psi_{[r,r-1,\ldots,r-(k-1)]}$ is $\delta_k=k(N-k)$.

\bigskip

Matrix elements of the group element $G$ in arbitrary $F_k$ are expressed through those
in $F$ by determinant formulas:
\be
g^{(k)}\left(\begin{array}{ccc}i_1 &\ldots & i_k \\ j_1 & \ldots & j_k\end{array}\right)
= \ \left<\psi_{[i_1}\ldots \psi_{i_k]}\Big| G \Big| \psi_{[j_1}\ldots \psi_{j_k]}\right> \
=\  \det_{1\leq a,b \leq k}\ G^{i_a}_{j_b}
\ee
where $G^a_b =\ <\psi_a|G|\psi_b>$.
In particular,
\be
{\cal G}_k=\ <\text{hw}_k| G |\text{hw}_k>\ = \det_{0\leq a,b\leq k-1} G^{a}_b
\ee
and
\be
{\cal G}_{-k} =
\ <-\text{hw}_k| G | \text{hw}_k>\ = \det_{0\leq a,b\leq k-1} G^{a}_{r-b}
\ee
The quantities $g^{(k)}$ satisfy a rich set of bilinear Pl\"ucker relations,
\be
g^{(k)}\left(\begin{array}{ccc}i_1 &\ldots & i_k \\ \Big[j_1 & \ldots & j_k\end{array}\right)
g^{(k')}\left(\begin{array}{cccc}i_1'&i_2' &\ldots & i_k' \\ j_{k+1}\Big]&j_1' & \ldots & j_{k-1}'
\end{array}\right)
=g^{(k+1)}\left(\begin{array}{cccc}i_1 &\ldots & i_k &\Big[i_{k'} \\ j_1 & \ldots & j_k&j_{k+1}
\end{array}\right)
g^{(k'-1)}\left(\begin{array}{ccc}i_1' &\ldots & i_{k-1}' \Big] \\
j_1' & \ldots & j_{k-1}'\end{array}\right)
\label{Plu}
\ee
which are often expressed as differential KP/Hirota like equations
for their generating functions called $\tau$-functions.

\bigskip

The standard $\tau$-function is defined as a "time-evolution" of ${\cal G}_{-k}$:
\be\boxed{
\tau^{(k)}(t,\bar t|G) = \
<\text{hw}_k| e^H\,G\, e^{\bar H} |\text{hw}_k>}
\ee
in $t,\bar t$ variables, which are associated with the operators $T^{(s)}_\pm$.
Namely, in accordance with (\ref{Tspowers}),
\be
e^H = \exp\left(\sum_{i=1}^k t_i R_k(T_+^i)\right) =
\left(\sum_{i=0}^k {\cal S}_iT_+^i\right)^{\otimes k}
\ee
and
\be
e^{\bar H} = \exp\left(\sum_{i=1}^k \bar t_i R_k(T_-^i)\right)
= \left(\sum_{i=0}^k \bar {\cal S}_iT_-^i\right)^{\otimes k}
\ee
where "coproducts"
\be
R_k(T_\pm^i) = T_\pm^i \otimes I \otimes \ldots \otimes I +
I\otimes T_\pm^i \otimes \ldots \otimes I + \ldots + I\otimes I\otimes \ldots \otimes T_\pm^i
\ee
and ${\cal S}_i={\cal S}_i(t)$ and $\bar {\cal S}_i = {\cal S}_i(\bar t)$
are the Schur polynomials of the time-variables $t$
and $\bar t$ respectively:
these are the Schur functions for pure symmetric representations satisfying
\be
\exp\left(\sum_{i=1}^\infty t_iz^i\right) = \sum_{i=0}^\infty {\cal S}_i(t) z^i
\ee
and
\be
\frac{\p {\cal S}_j(t)}{\p t_i} = \frac{\p^i {\cal S}_j(t)}{\p t_1^i} = {\cal S}_{j-i}(t)
\ee
As a corollary of these formulas,
the $\tau$-function has a simple determinant representation:
\be
\tau^{(k)}(t,\bar t,G) =
\sum_{{i_1,\ldots, i_k}\atop{j_1,\ldots,j_k}}
{\cal S}_{i_1}\ldots V_{i_k} \
\Big<\Psi_{[r-i_1,\ldots,r+1-k-i_k]}\Big|G\Big|\Psi_{[j_1,\ldots,k-1+j_k]}\Big>
\ \bar {\cal S}_{j_1}\ldots \bar {\cal S}_{j_k} \ =\
\det_{1\leq \alpha,\beta \leq k} {\cal H}^\alpha_\beta(t,\bar t,G)
\label{tauthroughH}
\ee
with
\be
{\cal H}^\alpha_\beta(t,\bar t,G) = \sum_{i,j} G^i_j\cdot {\cal S}_{i-\alpha}\bar {\cal S}_{j-\beta}
\ee
This ${\cal H}$ is independent of $k$
and it can be already considered as a function of infinitely many
$t$ and $\bar t$ time-variables, though for $SL(N)$ it is actually constrained by
\be
\frac{\p^N{\cal H}}{\p t_1^N} = \frac{\p^N{\cal H}}{\p \bar t_1^N} = 0 \nn \\
\ldots \nn \\
\frac{\p{\cal H}}{\p t_i} = \frac{\p{\cal H}}{\p\bar t_i} = 0 \ \ \ \text{for} \ \ i>N
\ee
It satisfies the characteristic shift relations
\be
\frac{\p}{\p t_i}{\cal H}^\alpha_\beta = {\cal H}^{\alpha+i}_\beta, \ \ \ \ \
\frac{\p}{\p \bar t_i}{\cal H}^\alpha_\beta = {\cal H}^{\alpha}_{\beta+i}
\ee
Exactly like those of $G$ itself, various minors of the matrix ${\cal H}$
satisfy the bilinear Pl\"ucker relations (\ref{Plu}), e.g.
$$
H\left(\begin{array}{ccc}1&\ldots & k \\ 1&\ldots & k\end{array}\right)
H\left(\begin{array}{cccc}k+1&1&\ldots & k-1 \\ k+1& 1&\ldots & k-1\end{array}\right) -
H\left(\begin{array}{cccc}1&\ldots & k-1&k \\ 1&\ldots & k-1&k+1\end{array}\right)
H\left(\begin{array}{cccc}k+1&1&\ldots & k-1 \\ k&1&\ldots & k-1\end{array}\right) =
$$
\vspace{-0.4cm}
\be
=
H\left(\begin{array}{ccc}1&\ldots & k+1 \\ 1&\ldots & k-1\end{array}\right)
H\left(\begin{array}{ccc}1&\ldots & k+1 \\ 1&\ldots & k-1\end{array}\right)
\ee
however, now the shift property
allows one to convert them into bilinear differential equations, of which the
simplest example is the Toda lattice relation
\be\label{todaeq}
\boxed{
\tau^{(k)}\p_1\bar\p_1 \tau^{(k)} - \p_1\tau^{(k)}\bar \p_1\tau^{(k)} =
\tau^{(k+1)}\tau^{(k-1)}
}
\ee
As emphasized in \cite{GKLMM,MMV}, if one defines the $\tau$-function with the help of
(\ref{tauthroughH}), but using another set of polynomials instead of the Schur ones,
one gets the same system of equations in another form:
e.g. for the $q$-Schur polynomials the Toda lattice equation will become difference.
This freedom leads to the notion of equivalent integrable hierarchies \cite{eqhier}.

\subsection{A simple exercise with Toda equations}

In the case under consideration, there is a "boundary" condition for the standard $\tau$-function $\tau_n$ as a function of $n$: $\tau_0=1$. Then, the Toda equation (\ref{todaeq}) can be used to recursively obtain any $\tau^{(k)}$ from $\tau^{(1)}$.

\bigskip

As an archetypical example
 consider the case of a single non-vanishing pair of time-variable: $t_1,\bar t_1\neq 0$.
Then for
\be
\tau^{(k)} = \ \Big<\text{hw}_k\Big|e^{tT_+}e^{\bar t T_-}\Big|\text{hw}_k\Big>\  =
\sum_{i=0}^{k(N-k)} \xi_i^{(k)}(t\bar t)^i
\ee
we have a recursion in $k$:
\be
\sum_{i,j}\xi_i^{(k+1)}\xi_j^{(k-1)}(t\bar t)^{i+j} =
-\sum_{i<j}(i-j)^2 \xi_i^{(k)}\xi_j^{(k)} (t\bar t)^{i+j-1}
\ee
The degrees in $t\bar t$ of both sides of this equation are $2Nk-k^2-1$.

For $\tau^{(1)} = (1+t\bar t)^r$ from (\ref{exatau1}) and $\tau^{(0)}=1$,
the Toda equation gives
\be
\tau^{(k)} =(1+t\bar t)^{k(r+1-k)}\cdot \prod_{j=1}^{k-1}
j!(r+1-k+j)^j
= (1+t\bar t)^{\delta_k} \cdot \prod_{j=0}^{k-1} \left<\psi_j\Big|\psi_j\right>=\nn\\
=  (1+t\bar t)^{k(N-k)} \cdot \left<\psi_{[0}\ldots \psi_{k-1]}\Big|
\psi_{[0}\ldots \psi_{k-1]}\right>
\ee
what is indeed the right answer for
$\ \left<\psi_{[0}\ldots \psi_{k-1]}\Big|e^{sT_+}e^{sT_-}\Big|\psi_{[0}\ldots \psi_{k-1]}\right>$.

In other words, in this simple example one has
\be
\tau^{(k)} \sim \Big(\tau^{(1)}\Big)^{\frac{k(N-k)}{N-1}}
\label{taukpower}
\ee
Note that in this form there is a limit at $N\rightarrow \infty$,
when $\tau^{(k)}\to \Big(\tau^{(1)}\Big)^k$ up to a factor. In order to deal with the factor, one first needs to rescale the time variables $t\to t/\sqrt{N}$, $\bar t\to\bar t/\sqrt{N}$ in order to have finite $\tau^{(1)}$ in the limit of $N\to \infty$. Then, $\tau^{(1)}=\exp (t\bar t)$ and
\be
\tau^{(k)}=\left(\prod_{i=1}^{k-1}i!\right) \times e^{kt\bar t}
\ee
in this limit.

\section{Skew $\tau$-function
\label{skewtaus}}

A skew $\tau$-function, where the right state is the lowest rather than the highest weight
is independent of conjugate time-variables $\bar t$, because $e^{\bar H}|-\text{hw}_k>\ = |-\text{hw}_k>$:
\be
\tau^{(k)}_-(t,G) = \ <\text{hw}_k| e^H\,G |-\text{hw}_k> \ =
\left(\frac{\p}{\p \bar t_1}\right)^{k(N-k)} \tau^{(k)}(s,\bar s, G)
\ee
At the first glance, this quantity is very far from possessing any nice properties:
the multiple differentiation destroys integrability structures, or at least,
hides them very deeply, and the results have hardly a finite limit of large $N$.
Surprisingly or not, this expectation is not quite true:
the skew $\tau$-functions turn out to be nice enough and deserve a separate study. We consider below the skew $\tau$-functions
with differently parameterized group element
\be\label{stau}\boxed{
\tau^{(k)}_-(w) = \ <\text{hw}_k| e^{\Lambda(w)} |-\text{hw}_k>}
\ee
with parameters (times) $w_k$  defining the matrix
\be
\Lambda (w) =   T_++ \sum_{s} w_s T_-^{(s)}
\label{Tps}
\ee
These parameters are related to $t$-variables in (\ref{skewtau}) by a non-linear transformation, and (\ref{stau})
has to be reduced yet to form (\ref{skewtau}) by the Gauss decomposition. This parametrization better suits the second section of this paper.

\subsection{The simplest example: a smell of integrability
\label{elem}}

Especially simple is the case when $G$ is made only from the $T_\pm$.
Despite $e^{T_++w^2T_-} \neq  e^{T_+}e^{w^2T_-}$, such a decomposition is easy to find (note that $w^2=w_1$ in (\ref{Tps})).
Indeed, since $T^\pm$ are generators of the $SL(2)$ subgroup in $SL(N)$,
one can just find the Gauss decomposition for $2\times 2$ matrices and then lift it
straightforwardly to arbitrary $N$:
{\footnotesize
\be
\!\!\!\!\!\!\!\!
\exp\!\left(\begin{array}{cc} 0& 1 \\ w^2 & 0\end{array}\right) =
\left(\begin{array}{cc} \cosh w  & \boxed{\frac{\sinh w}{w}} \\ \\ w\,\sinh w & \cosh w\end{array}\right)
= \left(\begin{array}{cc}  1 & \frac{\sinh w}{w\cdot\cosh w }\\ \\ 0&1    \end{array}\right)\!\!
\left(\begin{array}{cc}  \cosh^{-1}w & 0 \\ \\ 0 & \cosh w   \end{array}\right)\!\!
\left(\begin{array}{cc}  1 & 0 \\ \\ \frac{w\cdot \sinh w}{\cosh w} & 1   \end{array}\right)
\ee}
implies
{\footnotesize
\be
\exp\Big(T_++w^2T_-\Big)_{3\times 3} =
\exp\left(\begin{array}{ccc} 0&\sqrt{2}&0 \\ \\ w^2\sqrt{2} & 0& \sqrt{2} \\ \\
0&w^2\sqrt{2} & 0 \end{array}\right) =
\left(\begin{array}{ccc}
\cosh^2w  & \frac{\sqrt{2}\sinh w\cosh w}{w} & \boxed{\frac{\sinh^2w}{w^2}}\\ \\
w \,  {\sqrt{2}\sinh w\cosh w}  & 2\sinh^2+1 w & \frac{\sqrt{2}\sinh w\cosh w}{w} \\ \\
w^2\cdot\sinh^2(w)& w\,{\sqrt{2}\sinh w\cosh w} & \cosh^2(w)
\end{array}\right) = \nn \\ \nn \\ \nn \\
\!\!\!\!\!\!\!
= \exp\!\!\left(\begin{array}{ccc}  0 & \sqrt{2}\,\frac{\sinh(w)}{w\cdot\cosh w}& 0\\ \\
0& 0 & \sqrt{2}\,\frac{\sinh w}{w\cdot\cosh w} \\ \\ 0 & 0 & 0   \end{array}\right)\!\!
\cdot \!\left(\begin{array}{ccc}  \cosh^{-2}w & \\ \\ & 1 & \\ \\ & & \cosh^2w  \end{array}\right)\!\!
\cdot \exp\!\!\left(\begin{array}{ccc}  0 & 0 & 0\\ \\ \sqrt{2}\,\frac{w\cdot \sinh w}{\cosh w} & 0 & 0 \\ \\
0 & \sqrt{2}\,\frac{w\cdot \sinh w}{\cosh w} & 0
   \end{array}\right)
\ee

\bigskip

\be
\exp\Big(T_++w^2T_-\Big)_{4\times 4} =
\exp\left(\begin{array}{cccc} 0&\sqrt{3}&0& 0 \\ \\ w^2\sqrt{3} & 0& 2 & 0 \\ \\
0&2w^2 &0&\sqrt{3} \\ \\
0& 0 &w^2\sqrt{3}  & 0\end{array}\right) =\\=
\left(\begin{array}{cccc}
\cosh^3w  & \frac{\sqrt{3}\sinh w\cosh^2 w}{w} & \frac{\sqrt{3}\sinh^2 w\cosh w}{w^2} &  \boxed{\frac{\sinh^3w}{w^3}}\\ \\
w \,  \sqrt{3}\sinh w\cosh^2 w  & (3\sinh^2 w+1)\cosh w & \frac{(3\sinh^2w+2)\sinh w}{w}
&\frac{\sqrt{3}\sinh^2 w\cosh w}{w^2} \\ \\
w^2\,\sqrt{3}\sinh^2 w\cosh w&w\,(3\sinh^2w+2)\sinh w&(3\sinh^2 w+1)\cosh w&
\frac{\sqrt{3}\sinh w\cosh^2 w}{w}\\ \\
w^3\cdot\sinh^3(w)& w^2\,\sqrt{3}\sinh^2 w\cosh w  & w\,{\sqrt{3}\sinh w\cosh^2 w} & \cosh^3(w)
\end{array}\right) =
\ee

\bigskip

\centerline{
$
\!\!\!\!\!\!\!
= \exp\!\!\left(\begin{array}{cccc}  0 & \sqrt{3}\,\frac{\sinh(w)}{w\cdot\cosh w}& 0&0\\ \\
0&0&2\,\frac{\sinh(w)}{w\cdot\cosh w} & 0 \\ \\
0& 0 & 0& \sqrt{3}\,\frac{\sinh w}{w\cdot\cosh w} \\ \\ 0 & 0 & 0 & 0   \end{array}\right)\!\!
\cdot \!\left(\begin{array}{cccc}  \cosh^{-3}w & \\ \\ & \cosh^{-1}w & \\ \\
&&\cosh w \\ \\
& & & \cosh^2w  \end{array}\right)\!\!
\cdot \exp\!\!\left(\begin{array}{cccc}  0 & 0 & 0&0\\ \\ \sqrt{3}\,\frac{w\cdot \sinh w}{\cosh w} &0& 0&0 \\ \\
0&2\,\frac{w\cdot \sinh w}{\cosh w} &0& 0 \\ \\
0 & 0&\sqrt{3}\,\frac{w\cdot \sinh w}{\cosh w} & 0
   \end{array}\right)
$}}

\bigskip

\noindent
and so on.
Thus the first skew $\tau$-functions, the matrix elements in the {\it upper right} corner
of above matrices in this case are immediately calculated from the decomposition formula
in the last line:
\be
\tau^{(1)}_-\left(e^{T_++w^2T_-}\right)
= \left(e^{T_++w^2T_-}\right)_{1N} =
\frac{1}{r!}\cdot \left(\prod_{i=1}^r \sigma_i\right)\left(\frac{\sinh(w)}{w\cdot \cosh(w)}\right)^r
\cosh(w)^r = \left(\frac{\sinh(w)}{w}\right)^{N-1}
\label{tau1SL2}
\ee
Likewise by taking adjacent minors of the size $k$, one gets
\be
\tau^{(k)}_-\left(e^{T_++w^2T_-}\right)
= \det_{1\leq a,b\leq k} \left(e^{T_++w^2T_-}\right)_{a,N-b} =
\left(\frac{\sinh(w)}{w}\right)^{k(N-k)} =
\Big(\tau^{(1)}_-\Big)^{\frac{k(N-k)}{N-1}}
\label{taukSL2}
\ee
what is exactly the same relation as (\ref{taukpower}). Moreover, in order to have finite $\tau^{(1)}_-$ as $N$ goes to infinity, one has to rescale the time variables $w\to w/\sqrt{N}$ like it was in the case of the standard $\tau^{(1)}$ in the previous section.

Since $\sum_{k=1}^{N-1} k(N-k) = \frac{N(N^2-1)}{6}$, it follows that in this case
\be
\tau^{\rho}_-\left(e^{T_++w^2T_-}\right) =
\prod_{k=1}^r\tau^{(k)} = \left(\frac{\sinh(w)}{w}\right)^{\frac{N(N^2-1)}{6}}
\ee
and, hence, one would need a different rescaling to have a finite limit at infinite $N$.

\subsection{Importance of group structure}

Before we proceed further, it deserves making an important comment.
As explained in \cite{GKLMM} of crucial importance for integrability is
the choice of time variables.
Bilinear relations are at the very core of Lie/Hopf algebra theory
and they are immediately available for arbitrary groups,
classical or quantum, and families of their representations.
The problem is to convert these relations into equations
for generating functions, and much depends on the skill to build
the proper ones.
In fact, this is very well illustrated already by the simple example
of (\ref{tau1SL2}).
Imagine that instead of the clever choice of the
principally embedded $SL(2)$ generators $T_\pm$ we took just something "reasonable",
e.g.

\be
\exp\left(
\begin{array}{cccc}
 0 & \alpha_0  &0 & 0 \\
 \alpha_0 w^2 & 0 & \beta_0 & 0 \\
 0 & \beta_0 w^2 & 0 & \alpha_0 \\
 0 & 0 & \alpha_0w^2 & 0
 \end{array}\right) = \sum_{n=0}^\infty \frac{1}{(2n)!}
\left( \begin{array}{cccc}
  \xi_{n}w^{2n}  &0 & \eta_nw^{2n-2} & 0  \\
0 & \zeta_{n}w^{2n} & 0 & \eta_{n}w^{2n-2}  \\
  \eta_{n}w^{2n+2} & 0 & \zeta_{n}w^{2n} & 0\\
0& \eta_{n}w^{2n+2} & 0 & \xi_{n}w^{2n}
 \end{array}\right) + \nn \\ \nn \\
 +\sum_{n=0}^\infty \frac{1}{(2n+1)!}
\left(\begin{array}{cccc}
 0 & \alpha_{n}w^{2n+2}  &0 & \gamma_{n}w^{2n} \\
 \alpha_{n}w^{2n+4} & 0 & \beta_{n}w^{2n+2} & 0 \\
 0 & \beta_{n}w^{2n+4} & 0 & \alpha_{n}w^{2n+2} \\
 \gamma_{n}w^{2n+6} & 0 & \alpha_{n}w^{2n+4} & 0
 \end{array}\right)
\ee

\noindent
in the $4\times 4$ case.

Then, the evolution law for the $\alpha,\beta,\gamma$ parameters is
\be
\left(\begin{array}{c} \alpha_{n+1} \\ \beta_{n+1}\end{array}\right) =
\left(\begin{array}{cc} \alpha_0^2 & \alpha_0\beta_0 \\
\alpha_0\beta_0 & \alpha_0^2+\beta_0^2  \end{array}\right)
\left(\begin{array}{c} \alpha_{n} \\ \beta_{n}\end{array}\right) \nn \\
\nn \\
\gamma_{n+1} = \alpha_0\beta_0\alpha_n + \alpha_0^2\gamma_n
\ee
Symmetric shape of the matrices is preserved, because the evolution
possesses a conservation law
\be
\alpha_0^{-2n}\Big(\alpha_0\gamma_n + \beta_0\alpha_n-\alpha_0\beta_n\Big) = \text{constant}
\ee
On the other hand,
if this quantity was zero at $n=0$ (i.e. if $\gamma_0=0$), it remains zero for all $n$.
Then, the evolution of the corner element $\gamma$ simplifies:
\be
\gamma_{n+1} = \alpha_0^2\beta_n
\ee
and it remains to study only $\alpha,\beta$ sector.
The eigenvalues of the evolution operator  are ugly:
\be
\alpha_0^2+\frac{\beta_0^2}{2}\pm\beta_0\sqrt{\alpha_0^2+\frac{\beta_0^2}{4}}
\label{eigevo}
\ee
but they are already enough to distinguish between
the "naive" choice $\alpha_0=\beta_0=1$ and the "clever" ($SL(2)$) choice $\alpha=\sqrt{3},\beta=2$.
In these two cases, the eigenvalues we get respectively:
\be
\begin{array}{c|ccccccccccccc}
n & 0&1&2&3&4&5&6&7&8&9&\ldots   \\
\hline
\alpha_n & 1& 2&5&13&34&89&233&610&1597&4181&  \\
\beta_n  & 1&3&8&21&55&144&377&987&2584&6765&  \\
\hline
\gamma_n & 0 &   1&3&8&21&55&144&377&987&2584& \\
\frac{\gamma_n}{(2n+1)!} & 0 & \frac{1}{6}& \frac{1}{40}& \frac{1}{630} & \frac{1}{17280}
&\frac{1}{725760} & \frac{1}{43243200} & \frac{29}{ 100590336000  }
&\frac{47}{16937496576000} & \frac{1}{47076277248000}
\end{array}
\ee
and
\be
\begin{array}{l|ccccccccccccc}
n & 0&1&2&3&4&5&6&7&8&9&\ldots   \\
\hline
\alpha_n=\sqrt{3}\times & 1 &7&61&547&4921&44287&398581&3587227&32285041&290565367& \\
\beta_n     & 2& 20&182&1640&14762&132860&1195742&10761680&96855122&871696100&     \\
\hline
\gamma_n & 0 & 6&60&546&4920&44286&398580&3587226&32285949&290565366& \\
\frac{\gamma_n}{(2n+1)!} & 0 & 1&\frac{1}{2}&\frac{13}{120}&
\frac{41}{3024}&\frac{671}{604800}&\frac{73}{1140480}&\frac{597871}{217945728000}
&\frac{7913}{87178291200}&\frac{28009}{11725959168000}&
\end{array}
\ee
The sequences of $\gamma_n$ seem equally ugly in the both cases, but in fact in the latter case
$\frac{\gamma_n}{(2n+1)!}$ are exactly expansion coefficients of $\sinh^3(x)$, while they represent
no nice function in the former case.
This is of course clear from looking at the eigenvalues (\ref{eigevo}):
in the latter case $\gamma_n = \frac{3}{4}(3^{2n}-1^{2n})$ and
$\sum_n \frac{\gamma_nx^{2n+1}}{(2n+1)!} = \frac{1}{4}\Big(\sinh{3x}-3\sinh(x)\Big)=\sinh^3(x)$,
while in the former case one rather gets $\gamma_n=\frac{1}{2^n\sqrt{5}}
\Big((1+\sqrt{5})^{2n}-(1-\sqrt{5})^{2n}\Big)$
and $\sum_n \frac{\gamma_nx^{2n+1}}{(2n+1)!} =
\frac{2}{\sqrt{5}}\Big(\frac{\sinh\frac{(1+\sqrt{5})x}{2}}{1+\sqrt{5}}
-\frac{\sinh\frac{(1-\sqrt{5})x}{2}}{1-\sqrt{5}}\Big)=\sinh\frac{x}{2}\cosh\frac{\sqrt{5}x}{2}
- \frac{1}{\sqrt{5}}\cosh\frac{x}{2}\sinh\frac{\sqrt{5}x}{2}$.
Clearly the $\sinh^3$ formula is easily continued to arbitrary $N$, while the other expression
has low chances for this.
The reason for this lies rather deep in group theory (not so deep in this particular case,
but deeper in more complicated situations), and is not revealed and put under control
in full generality.

\bigskip

Thus, even this trivial example can serve as a non-trivial illustration to the general formulation of
the $\tau$-function puzzle in \cite{GKLMM}.

\subsection{Expression through eigenvalues }

Coming back to our main line, once we suspect some integrability to be present,
the next natural thing to do is to look for matrix model representations,
where at the core of integrability lies just the Vandermonde
determinant.
Not surprisingly, such representation is immediately available:
\be
\tau^{(1)}_-= \left(\frac{\sinh(w)}{w}\right)^r
= \sum_{i=0}^r \frac{(-)^i\,r!}{i!(r-i)!}\cdot \frac{e^{(r-2i)w}}{(2w)^r}
= r!\cdot \sum_{i=1}^N  \frac{e^{\lambda_i}}{\prod_{j\neq i}(\lambda_i-\lambda_j)} \nn \\
\tau^{(2)}_-= \left(\frac{\sinh(w)}{w}\right)^{2(r-1)}
= -\frac{r!(r-1)!}{2!}\cdot \!\!\!\!
\sum_{i_1,i_2 =1}^N  \frac{(\lambda_{i_1}-\lambda_{i_2})^2\cdot e^{\lambda_{i_1}+\lambda_{i_2}}}
{\prod_{j\neq i_1}(\lambda_{i_1}-\lambda_{j})\prod_{j\neq i_2}(\lambda_{i_2}-\lambda_{j})}\nn \\
\nn\\ \ldots \nn \\ \nn \\
\tau^{(k)}_-= \left(\frac{\sinh(w)}{w}\right)^{k(r+1-k)}\!\!\!\!\!\!\!\!
= (-)^{k(k-1)/2}\cdot\frac{r!(r-1)!\ldots(r+1-k)!}{2!\ldots k!}\cdot\!\!\!\!\!\!\!\!
\sum_{i_1,\ldots,i_k =1}^N  \frac{\prod_{a<b}^k(\lambda_{i_a}-\lambda_{i_b})^2
\cdot e^{\lambda_{i_1}+\ldots+\lambda_{i_k}}}
{\prod_{j\neq i_1}(\lambda_{i_1}-\lambda_{j})\ldots\prod_{j\neq i_k}(\lambda_{i_k}-\lambda_{j})}\nn \\
\ldots
\label{tau1m}
\ee
where $\ \lambda_i = (N+1-2i)\cdot w \ $
are eigenvalues of the matrix $\Lambda={T_++w^2T_-}$.
This expression  through eigenvalues can seem unnecessary complicated,
however, it essentially says that
\be
\tau^{(k)} \sim \Delta_{k}^2=\prod_{a<b}^k(\lambda_{i_a}-\lambda_{i_b})^2,
\ee
which allows us to immediately recognize
a Toda-chain $\tau$-function, see s.\ref{skmamo} below.

\bigskip

Moreover, such a representation remains true for more general matrices (\ref{Tps}), \cite{HKP}
\be
\Lambda =   T_++ \sum_{s} w_s T_-^{(s)}
\ee
This is despite particular eigenvalues $\lambda_i$ are now incalculable roots of
polynomials of degree $N$: since the r.h.s. of (\ref{tau1m})
is actually made from symmetric polynomials of $\lambda_i$,
one can apply the Vieta formulas. In particular, they satisfy the condition
\be\label{al}
\sum_{i=1}^N\lambda_i=0
\ee

\subsection{Vandermonde-diagonalizable case}

We begin the study of relations (\ref{tau1m}) from a slightly simplified case.
Since we are interested in formulas expressed through eigenvalues of the matrix $\Lambda$,
it is reasonable to substitute
$\Lambda=V\cdot \text{diag}(\lambda_1,\ldots ,\lambda_N)\cdot V^{-1}$
\be
\tau^{(1)}_-\Big(e^\Lambda\Big) = \sum_{i=1}^N V_{1i}V^{-1}_{iN} \cdot e^{\lambda_i} \nn \\
\tau^{(2)}_-\Big(e^\Lambda\Big) = \sum_{i_1,i_2=1}^N \Big(V_{1i_1}V_{2i_2}-V_{2i_1}V_{1i_2}\Big)
V^{-1}_{i_1,N}V^{-1}_{i_2,N-1}
\cdot e^{\lambda_{i_1}+\lambda_{i_2}} \nn \\
\ldots
\ee
Then (\ref{tau1m}) turns into $\lambda$-dependent conditions on the matrix $V$:
\be
V_{1i}V^{-1}_{iN}\cdot \prod_{j\neq i}(\lambda_i-\lambda_j) =1\nn \\
\frac{1}{2!}\left\{\Big(V_{1i_1}V_{2i_2}-V_{2i_1}V_{1i_2}\Big)V^{-1}_{i_1,N}V^{-1}_{i_2,N-1}
+ (i_1\leftrightarrow i_2)\right\} \cdot
\prod_{j\neq i_1}(\lambda_{i_1}-\lambda_j)\prod_{j\neq i_2}(\lambda_{i_2}-\lambda_j)
=\frac{1}{2!}(\lambda_{i_1}-\lambda_{i_2})^2 \nn \\
\frac{1}{3!}\left\{\Big(\det_{a,b=1,2,3}V_{ai_b} \Big)V^{-1}_{i_1,N}V^{-1}_{i_2,N-1}V^{-1}_{i_3,N-2}
+ 5 \text{\ perms}(i_1,i_2,i_3\right\} \cdot
\prod_{a=1}^3 \prod_{j\neq i_a}(\lambda_{i_q}-\lambda_j)
=\frac{1}{3!}\,\prod_{a<b}^3(\lambda_{i_a}-\lambda_{i_b})^2 \nn \\
\ldots
\label{Uconds1}
\ee
The above mentioned simplification is that
we slightly changed the normalization factors at the r.h.s.,
so that the system (\ref{Uconds1}) is solved by the Vandermonde matrix
\be
V=\tilde V = \{\lambda_j^{i-1}\ | \ i,j=1,...,N\}
\ee
i.e. the story applies to the constant connection
\be
\tilde\Lambda=\tilde V\cdot \text{diag}(\lambda_1,\ldots ,\lambda_N)\cdot \tilde V^{-1}
= \tilde T_+ + \sum_{s=1}^{r+1}(-)^{s+1}
\left(\sum_{i_1,\ldots,i_s}\prod_{i_1<\ldots<i_s} \lambda_{i_1}\ldots \lambda_{i_s}\right) E_{r,r+1-s}
= \nn \\  \nn \\
= \left(\begin{array}{ccccccccc}
0&1&0&\ldots& 0 \\ 0&0&1&&0 \\ \ldots \\ 0&0&0&\ldots &1 \\
(-)^r\prod_i \lambda_i\  &\ (-)^{r-1}\prod_i\lambda_i \sum_i \lambda_i^{-1}\
&\ (-)^r\prod_i\lambda_i \sum_{i<j}(\lambda_i\lambda_j)^{-1}\ &&\ \sum_i \lambda_i
\end{array}
\right)
= \nn \\ \nn \\
= \tilde T_+ + \sum_{s=1}^{r+1}(-)^{s+1} {\cal S}_{[1^s]}(\lambda)\cdot E_{r,r+1-s}
\label{Ttp}
\ee
Since the diagonalizing matrix for (\ref{Ttp}) is $V_{ij} = \lambda_j^{i-1}$,
the eigenvector of (\ref{Ttp}) with eigenvalue $\lambda_j$ is $v^{(j)}_i = \lambda_j^{i}$.
Since (\ref{Ttp}) is not symmetric, the eigenvectors and the diagonalizing matrix are not orthogonal.

Note that $\tilde T_+$ in (\ref{Ttp})
\be
\tilde T_+ = \left(\begin{array}{ccccc} 0&1&0&\ldots& 0 \\ 0&0&1&&0 \\ \ldots \\ 0&0&0&&1 \\ 0&0&0&&0
\end{array}\right) = \sum_{i=1}^r    T_{\alpha_i}
\ \ \ \boxed{\neq} \ \ \ \ \sum_{i=1}^r \sqrt{i(r+1-i)}\cdot T_{\alpha_i} = T_+
\ee
Thus there are two differences between (\ref{Ttp}) and (\ref{Tps}):
$\tilde T_+\neq T_+$ and relations (\ref{Uconds1})
differ from (\ref{tau1m}) by the lack of some factorials.
Remarkably, these differences compensate each other.
We provide an exhaustive explanation for this in subsection \ref{ltr} below,
but before that it deserves looking closer at (\ref{Ttp}).

\subsection{Schur expansions
\label{skmamo}}

In (\ref{Ttp}) ${\cal S}_R(\lambda)$ are Schur symmetric functions of variables $\{\lambda_i\}$,
and only pure antisymmetric (i.e. fundamental) representations $R=[1^k]$
contribute to $\tilde\Lambda$.
The corner element of this $e^{\tilde\Lambda}$ is especially simple:
\be
\left(e^{\tilde\Lambda}\right)_{1N}=\sum_k^\infty {{\cal S}_{[k]}(\lambda)\over (k+r)!}
\label{teLcorner}
\ee
where the (infinite) sum is now over all {\it pure symmetric} representations $R=[k]$.
Since this class of Schur functions is generated by
\be
\sum_{k=0}^\infty z^k {\cal S}_{[k]} = \exp\left(\sum_{k=0}^\infty \frac{p_kz^k}{k}\right)
= \prod_{i=1}^N (1-z\lambda_i)^{-1}
\ee
the sum in (\ref{teLcorner}) is obtained by the Laplace/Pade transform:
\be
\left(e^{\tilde\Lambda}\right)_{1N}=\sum_k^\infty {{\cal S}_{[k]}(\lambda)\over \Gamma(k+r+1)} =
\frac{1}{\pi}\sum_k^\infty (-)^k   \Gamma(-k-r){\cal S}_{[k]}(\lambda)
= \frac{1}{\pi}\int_0^\infty \frac{e^{-x}dx}{x^{N}} \sum_k \frac{{\cal S}_{[k]}}{(-x)^k} = \nn \\
= \frac{1}{\pi}\int_0^\infty \frac{e^{-x} dx}{\prod_{j=1}^N (x+\lambda_j)}
\ \longrightarrow \ \sum_{i}\frac{e^{\lambda_i}}{\prod_{j\neq i}(\lambda_i-\lambda_j)}
\ee
if the integral is substituted by the sum over residues at $x=-\lambda_i$.

This explains the first of peculiar formulas in (\ref{tau1m}).
The simplest way to understand what happens for higher fundamental representations is to look at
the $r=1$ case, then the entire matrix
\be
e^{\tilde\Lambda} =
\exp\left(\begin{array}{cc}
0 & 1 \\ -{\cal S}_{[11]} & {\cal S}_{[1]}
\end{array}\right) =
\left(\begin{array}{cc}
1 - {\cal S}_{[11]}\sum_{k=0}^\infty \frac{{\cal S}_{[k]}}{(k+2)!}
& \sum_{k=0}^\infty \frac{{\cal S}_{[k]}}{(k+1)!}
\\ \\
-{\cal S}_{[11]}\sum_{k=0}^\infty \frac{{\cal S}_{[k]}}{(k+1)!}
& \sum_{k=0}^\infty \frac{{\cal S}_{[k]}}{k!}
\end{array}\right)
\label{etLN2}
\ee
is $2\times 2$ and its determinant is a bilinear sum, giving rise to the double integral
\be
{\cal S}_{[11]}\cdot \sum_{n_1,n_2} \left(\frac{1}{(n_1+r+1)!(n_2+r-1)!}-\frac{1}{(n_1+r)!(n_2+r)!}\right)
{\cal S}_{[n_1]}{\cal S}_{[n_2]} = \nn \\
= \frac{{\cal S}_{[11]}}{\pi^2}
\int\int \frac{e^{-x-y}\,dxdy}{\prod_{j=1}^N (x+\lambda_j)(y+\lambda_j)}\left(\frac{x}{y}+\frac{y}{x}-2\right)
\label{doublint}
\ee
where the last factor comes from the deviations of the factorial argument shifts from $r$
and symmetrization w.r.t. $x\leftrightarrow y$.
Switching to the residue sums, one obtains (for this particular case) the second formula in (\ref{tau1m})
\be
\sum_{i_1,i_2} \frac{(\lambda_{i_1}-\lambda_{i_2})^2\cdot
e^{\lambda_{i_1}+\lambda_{i_2}}}{\prod_{j\neq i_1}(\lambda_{i_1}-\lambda_j)
\prod_{j\neq i_2}(\lambda_{i_2}-\lambda_j)}
\ee
For arbitrary $N=r+1$ eq.(\ref{teLcorner}) is raised to
\be
\left(e^{\tilde\Lambda}\right)_{n,N-m}=(-)^m
\sum_k^\infty {{\cal S}_{[k+n-1,1^m]}(\lambda)\over (k+r)!}
\label{teL}
\ee
Eq.(\ref{etLN2})  for $r=1$ arises from this if one uses the recursions
\be
{\cal S}_{[km]} \ \stackrel{N=2}{=}\ {\cal S}_{[11]}^m {\cal S}_{[k-m]} \nn \\
{\cal S}_{[k+1]}\ \stackrel{N=2}{=}\ {\cal S}_{[k]}{\cal S}_{[1]}-{\cal S}_{[k-1]}{\cal S}_{[11]}
\ee

\subsection{Matrix model representations for $\tau_-(e^{\tilde\Lambda})$ and integrable properties of the skew $\tau$-function}

It remains to describe the differential equations (integrable hierarchy)
which our skew tau-functions satisfy.

As a straightforward generalization of (\ref{doublint}),
 we obtain the skew $\tau$ function for the "eigenvalue-inspired"
constant connection (\ref{Tpsev1}) expressed
by the multiple integral
\be\label{mamo}
\tau^{(k)}_-(e^{\tilde\Lambda}) =
\int \Delta(x)^2 \prod_{a=1}^k \frac{e^{-tx_a}}{\prod_{i=1}^N (x_a+\lambda_i)}{dx_a\over\pi}
\ee
i.e. one can identify it with the Hermitian matrix model with a logarithmic potential.

Introduction of  $t$ is
equivalent to rescaling $w_s \longrightarrow w_st^s$  and
$\tau^{(k)}_- \longrightarrow t^{k(N-k)}\cdot \tau^{(k)}_-$.
Alternatively, one can substitute $t$-derivative by action of the dilatation operator
$\sum_i \lambda_i\frac{\p}{\p\lambda_i}$

As known from \cite{GMMMO,UFN23}, the matrix integrals of kind (\ref{mamo}) satisfy the Toda chain equation in the
variable $t$.
This is immediately obvious already from (\ref{tau1m}).
The action of $t$-derivative on $\tau^{(k)}_-$ of this form provides $\sum_a \lambda_i$,
then, say,
\be
\tau^{(1)}_-\,\frac{\p^2\tau^{(1)}_-}{\p t^2}\ -\ \left(\frac{\p\tau^{(1)}_-}{\p t}\right)^2
= \sum_{i_1,i_2} \frac{e^{t(\lambda_{i_1}+\lambda_{i_2})}}{\prod_{j\neq i_1}(\lambda_{i_1}-\lambda_j)
\prod_{j\neq i_2}(\lambda_{i_2}-\lambda_j)}\cdot (\lambda_{i_2}^2-\lambda_{i_1}\lambda_{i_2})
=  \tau^{(2)}_-
\ee
because the symmetry between $\lambda_{i_1}$ and $\lambda_{i_2}$
allows one to substitute the last factor   by $\frac{1}{2!}(\lambda_{i_1}-\lambda_{i_2})^2$.
Similarly, acting on $\tau^{(2)}_-$, one gets in the numerator
\be
\frac{1}{2!\cdot 2!}\,(\lambda_{i_1}-\lambda_{i_2})^2(\lambda_{i_3}-\lambda_{i_4})^2\cdot
\Big((\lambda_{i_3}+\lambda_{i_4})^2
- (\lambda_{i_1}+\lambda_{i_2})(\lambda_{i_3}+\lambda_{i_4})\Big)
\ee
which is equivalent to
\be
\frac{1}{3!}\, (\lambda_{i_1}-\lambda_{i_2})^2(\lambda_{i_1}-\lambda_{i_3})^2(\lambda_{i_2}-\lambda_{i_3})^2
\ee
if the both quantities are totally symmetrized.
In the same way, one gets for arbitrary $k$:
\be\boxed{
\tau^{(k)}_-\,\frac{\p^2\tau^{(k)}_-}{\p t^2}\ -\ \left(\frac{\p\tau^{(k)}_-}{\p t}\right)^2
 =  \tau^{(k+1)}_-\,\tau^{(k-1)}_-}
\ee

Note that the matrix integral (\ref{mamo}) has the determinant representation of the typical Toda chain form:
\be
\tau_-^{(k)}=\det_{1\le i,j\le k} C_{i+j-2}
\ee
with
\be
C_k=\sum_{i=1}^N \frac{\lambda_i^k
e^{\lambda_i}}{\prod_{j\neq i}(\lambda_i-\lambda_j)} =
\sum_{n}^\infty \frac{{\cal S}_{[n]}}{(n+r-k)!}
\ee
and
\be\label{Todam}
\tau_-^{(0)}=1,\ \ \ \ \ \tau_-^{(N+1)}=0
\ee
what means that it is a Toda molecule $\tau$-function, see details in \cite{Mtau}. In the limit of $N\to\infty$, it becomes the full Toda forced hierarchy \cite{KMMOZ}. In fact, in this limit, one would better look at another form of the matrix integral (\ref{mamo}) (at $t=1$):
\be\label{mamo1}
\tau^{(k)}_-(e^{\tilde\Lambda}) =\left({1\over\prod_i\lambda_i}\right)^k
\int \Delta(x)^2 \prod_{a=1}^k e^{-x_a+\sum_{m=1}^\infty t_mx_a^m}{dx_a\over\pi}
\ee
where the variables $\lambda_i$ play the role of Miwa variables related with the time variables $t_m$ as
\be
t_m={1\over m}\sum_{i=1}^N(-\lambda_i)^{-m}
\ee
At finite $N$ only finitely many ($N$) time variables $t_m$ are independent, while in the limit of $N\to \infty$ (\ref{mamo}) becomes, up to a factor of $\Big(\prod_i\lambda_i^{-1}\Big)^k$ a full-fledged $\tau$-function of the Toda chain hierarchy in all times $t_m$.

\subsection{Diagonalizing (\ref{Tps})}

Let us now return to our original (\ref{Tps}).
Our next goal is to express   matrices (\ref{Tps}) through their eigenvalues: this should be possible,
because the number of parameters $w_s$
is exactly $r$, which is the number of independent eigenvalues of a traceless matrix of size $r+1$.
In what follows, $\sum_{i=1}^N \lambda_i = 0$.

In the simplest $2\times 2$ ($r=1$) case there is no difference between (\ref{Tps}) and (\ref{Ttp}):
\be
T_+ \underbrace{- \lambda_1\lambda_2}_{w^2=\lambda_1^2}\cdot T_- =
\left(\begin{array}{cc} 0 & 1 \\ w^2 & 0\end{array}\right)=
\left(\begin{array}{cc} 0 & 1 \\ -\lambda_1\lambda_2 & \lambda_1+\lambda_2 \end{array}\right)
= \left(\begin{array}{cc} 1 & 1 \\ \lambda_1&\lambda_2\end{array}\right)
\left(\begin{array}{cc} \lambda_1 \\ & \lambda_2 \end{array}\right)
\left(\begin{array}{cc} 1 & 1 \\ \lambda_1&\lambda_2\end{array}\right)^{-1}
\label{Tps2}
\ee
Accordingly, the only factorial $r!=1$.

However, already for $3\times 3$ ($r=2$)   a difference shows up:
\be
T_+ \underbrace{- \frac{\lambda_1\lambda_2+\lambda_2\lambda_3+\lambda_1\lambda_3}{4}}_{w_1}\cdot T_-
+\underbrace{\frac{\lambda_1\lambda_2\lambda_3}{4}}_{w_2}\cdot T^{(2)}_- =
\left(\begin{array}{ccc} 0 & \sqrt{2} & 0 \\  \sqrt{2}w_1 &0 &\sqrt{2}\\
2w_2 & \sqrt{2} w_1& 0 \end{array}\right)=\nn\\
= \underbrace{\left(\begin{array}{ccc} 4 & 4 & 4  \\ 2\sqrt{2}\lambda_1&2\sqrt{2}\lambda_2 & 2\sqrt{2}\lambda_3\\
\lambda_1^2+\lambda_2\lambda_3 & \lambda_2^2+\lambda_1\lambda_3& \lambda_3^2+\lambda_1\lambda_2
\end{array}\right)}_{V}
\left(\begin{array}{ccc} \lambda_1 \\ & \lambda_2 \\ && \lambda_3 \end{array}\right)
\left(\begin{array}{ccc} 4 & 4 & 4  \\ 2\sqrt{2}\lambda_1&2\sqrt{2}\lambda_2 & 2\sqrt{2}\lambda_3\\
\lambda_1^2+\lambda_2\lambda_3 & \lambda_2^2+\lambda_1\lambda_3& \lambda_3^2+\lambda_1\lambda_2
\end{array}\right)^{-1}
\label{Tps3}
\ee
Now, we get (\ref{Uconds1}) with missing factorials restored, exactly what is needed for (\ref{tau1m}):
\be
V_{1i}V^{-1}_{iN}\cdot \prod_{j\neq i}(\lambda_i-\lambda_j) =\boxed{2!} \nn \\
\frac{1}{2!}\left\{\Big(V_{1i_1}V_{2i_2}-V_{2i_1}V_{1i_2}\Big)V^{-1}_{i_1,N}V^{-1}_{i_2,N-1}
+ (i_1\leftrightarrow i_2)\right\} \cdot
\prod_{j\neq i_1}(\lambda_{i_1}-\lambda_j)\prod_{j\neq i_2}(\lambda_{i_2}-\lambda_j)
=\frac{\boxed{2!}}{2!}\cdot(\lambda_{i_1}-\lambda_{i_2})^2 \nn \\
\frac{1}{3!}\left\{\Big(\det_{a,b=1,2,3}V_{ai_b} \Big)V^{-1}_{i_1,N}V^{-1}_{i_2,N-1}V^{-1}_{i_3,N-2}
+ 5 \text{\ perms}(i_1,i_2,i_3\right\} \cdot
\prod_{a=1}^3 \prod_{j\neq i_a}(\lambda_{i_q}-\lambda_j)
=\frac{\boxed{2!}}{\boxed{2!}\cdot 3!} \,\prod_{a<b}^3(\lambda_{i_a}-\lambda_{i_b})^2
\label{Vma3}
\ee

Similarly, for $4\times 4$ ($r=3$) case:

\bigskip

{\footnotesize
\be\label{4}
T_+ \underbrace{- \frac{1}{10}\left(\sum_{ i<j }^4\lambda_i\lambda_j\right) }_{w_1}\cdot T_-
+\underbrace{\frac{1}{24}\left(\sum_{ i<j<k}^4\lambda_i\lambda_j\lambda_k\right)}_{w_2}\cdot T^{(2)}_-
\underbrace{-\frac{1}{36}\left\{\lambda_1\lambda_2\lambda_3\lambda_4-\frac{9}{100}
\left(\sum_{i<j}^4\lambda_i\lambda_j\right)^2\right\}
}_{w_3}\cdot T^{(3)}_- =
\nn \\ =
\left(\begin{array}{cccc} 0 & \sqrt{3} & 0 & 0 \\  \sqrt{3}w_1 &0 &2 &0\\
2\sqrt{3}w_2 & 2 w_1& 0 & \sqrt{3} \\
6w_3 &  2\sqrt{3}w_2 & \sqrt{3}w_1& 0 \end{array}\right)=\nn\\
= \underbrace{
\left(\begin{array}{ccc} 60 &\ldots & 60\\ 20\sqrt{3}\lambda_1&& 20\sqrt{3}\lambda_4 \\
\frac{\sqrt{3}}{2}\Big(20\lambda_1^2 -3(\lambda_1^2+\lambda_2^2+\lambda_3^2+\lambda_4^2)\Big)
&&\frac{\sqrt{3}}{2}\Big(20\lambda_4^2 -3(\lambda_1^2+\lambda_2^2+\lambda_3^2+\lambda_4^2)\Big)\\
4\lambda_1^3- \lambda_1(\lambda_2^2+\lambda_3^2+\lambda_4^2) - 5\lambda_2\lambda_3\lambda_4
&\ldots &4\lambda_4^3- \lambda_4(\lambda_1^2+\lambda_2^2+\lambda_3^2) - 5\lambda_1\lambda_2\lambda_3
\end{array}\right)}_{V}
\left(\begin{array}{cccc} \lambda_1 \\ & \lambda_2 \\ && \lambda_3\\ &&&\lambda_4 \end{array}\right)
V^{-1}
\label{Tps4}
\ee}
and
{\footnotesize
\be
V_{1i}V^{-1}_{iN}\cdot \prod_{j\neq i}(\lambda_i-\lambda_j) =\boxed{3!} \nn \\
\frac{1}{2!}\left\{\Big(V_{1i_1}V_{2i_2}-V_{2i_1}V_{1i_2}\Big)V^{-1}_{i_1N}V^{-1}_{i_2,N-1}
+ (i_1\leftrightarrow i_2)\right\} \cdot
\prod_{j\neq i_1}(\lambda_{i_1}-\lambda_j)\prod_{j\neq i_2}(\lambda_{i_2}-\lambda_j)
=\frac{\boxed{3! \cdot 2!}}{2!}\cdot(\lambda_{i_1}-\lambda_{i_2})^2 \nn \\
\frac{1}{3!}\left\{\Big(\det_{a,b=1,2,3}V_{ai_b} \Big)V^{-1}_{i_1N}V^{-1}_{i_2,N-1}V^{-1}_{i_3,N-2}
+ 5 \text{\ perms}(i_1,i_2,i_3\right\} \cdot
\prod_{a=1}^3 \prod_{j\neq i_a}(\lambda_{i_q}-\lambda_j)
=\frac{\boxed{3!\cdot 2!}}{\boxed{2!}\cdot 3!}\,\prod_{a<b}^3(\lambda_{i_a}-\lambda_{i_b})^2 \nn \\
\frac{1}{4!}\left\{\Big(\det_{a,b=1,2,3,4}V_{ai_b} \Big)V^{-1}_{i_1N}V^{-1}_{i_2,N-1}V^{-1}_{i_3,N-2}
V^{-1}_{i_4,N-3}
+ 23 \text{\ perms}(i_1,i_2,i_3,i_4\right\} \cdot
\prod_{a=1}^4 \prod_{j\neq i_a}(\lambda_{i_q}-\lambda_j)
=\frac{\boxed{3!\cdot 2!}}{\boxed{2!\cdot 3!}\cdot 4!}\,\prod_{a<b}^4(\lambda_{i_a}-\lambda_{i_b})^2 \nn \\
\ldots
\label{Vma4}
\ee}
and so on.

\bigskip

In the $5\times 5$ ($r=4$) case
{\footnotesize
\be
\!\!\!\!\!\!\!\!\!\!\!\!\!\!\!\!\!\!\!\!
T_+ \underbrace{- \frac{1}{20}\left(\sum_{ i<j }^5\lambda_i\lambda_j\right) }_{w_1}\cdot T_-
+\underbrace{\frac{1}{84}\left(\sum_{ i<j<k}^5\lambda_i\lambda_j\lambda_k\right)}_{w_2}\cdot T^{(2)}_-
\underbrace{-\frac{1}{288}\left(\sum_{ i<j<k<l}^5\lambda_i\lambda_j\lambda_k\lambda_l-64w_1^4\right)}_{w_3}
\cdot T^{(3)}_-
+ \underbrace{\frac{1}{576}\left(\prod_{i=1}^5\lambda_i + 192 w_1^2w_2^2\right)}_{w_4}\cdot T^{(4)}
\label{Tpsev5}
\ee}

In general, the coefficients of the characteristic polynomial $\det_{N\times N}(zI+\Lambda)$
are:

\bigskip

\bigskip

\centerline{\footnotesize
$
\begin{array}{c|ccccccccccc}
r&w_1&w_2&w_3-w_1^2&w_4-w_2w_1& w_5-w_3w_1-w_2^2+w_1^3& w_6-w_4w_1-w_3w_2+w_2w_1^2
&\ldots
\\
\hline
1&1 \\
2&4&4   \\
3&10&24& 36-9 \\
4&20&84&288-64&  576-192 \\
5&35&224&1296-259& 5760-1760& 1440-3600-1600+225\\
6&56&504&4320-784& 31680-8928 & 172800 - 40320-18000+2304& 518400-103680-86400+13824\\
7&84 & 1008 &11880-1974 & 126720 - 33120 & 1123200 -245520-110016+12916
& 7257600-1370880-1149120+170352
&\ldots
\\
8&120&1848 & 28512-4368&411840-100320& 5241600-1077120-48176+52480
&54432000-9734400-8201088+1132032
&\ldots
\\
9& 165& 3168& 61776-8778& \\
10 &220 & 5148&123552-16368 & \\
11&286 &8008 &231660-28743 &
\\
& \ldots   \\
\hline
& \\
r &   \frac{N(N^2-1)}{6} &\frac{N(N^2-1)(N^2-2^2)}{5\cdot 6} &  \ldots
\end{array}
$
}

\bigskip

\bigskip

\noindent
The coefficient in front of $w_s$ is just
\be
\frac{(s!)^2}{(2s+1)!}\cdot \prod_{i=-s}^s (N+i)
\ee
This  means that (\ref{Tps}) will be
\be
\Lambda =
T_+ +\sum_{s=1}^{r=N-1}\left(\underbrace{(-)^s\, \frac{(2s+1)!}{(s!)^2\cdot \prod_{i=-s}^s (N+i)}
\sum_{i_1<\ldots< i_{s+1}}^N \lambda_{i_1}\ldots \lambda_{i_{s+1}} }_{w_s}\ + \ O(w^4)\right)
\cdot T_-^{(s)}
\label{Tpsev1}
\ee
The coefficients in front of $w_1^2$, $w_2w_1$, $w_3w_1$, $\ldots$ are
\be
\frac{r(r^2-1)(r^2-2^2)(5r+12)}{360}, \ \ \
\frac{r(r^2-1)(r^2-2^2)(r^2-3^2)(7r+20)}{1260}, \ \ \
\frac{r(r^2-1)(r^2-2^2)(r^2-3^2)(r^2-4^2)(9r+30)}{7560}, \ \ \
\ldots
\nn
\ee
i.e.
$w_{s-2}w_1$ (subtracted from $w_s$) with $k\geq 2$ ($s\geq 4$) comes with the coefficient
\be
\frac{(s-2)!(s-1)!}{3\cdot (2s-1)!}\cdot \Big((2s-1)r+s(s+1)\Big)\prod_{i=1-s}^{s-1}(r+i)
\ee
and for $w_1^2$ (subtracted from $w_3$) there is an extra factor of $\frac{1}{2}$.

More systematically, these seemingly complicated formulas can be obtained from
\be
\log\det_{N\times N}(I+z^{-1}\Lambda) = -\sum_{k=1}^\infty \frac{(-)^k}{kz^k}\Tr\Lambda^k
= -\sum_{k=1}^\infty \frac{p_k}{kz^k}
\label{logdet}
\ee
with the time-variables $p_k = \sum_{i=1}^N \lambda_i^k$.
Equation (\ref{logdet}) expresses these variables through $w_s$
with coefficients, made from
\be
\Xi_k(N)=\sum_{i=1}^{r+1-k} \sigma_i^2\sigma_{i+1}^2\ldots\sigma_{i+k-1}^2 =
\sum_{i=1}^{r+1-k} \frac{(i+k-1)!\,(r+1-i)!}{(i-1)!\,(r+1-i-k)!} =
\frac{(k!)^2}{(2k+1)!} \prod_{i=-k}^k (N+i)
\ee
and similar (more complicated) sums.

\bigskip

{\bf
Now the question is: \ what is so special about these two families
(\ref{Ttp}) and (\ref{Tps})?
}

\bigskip

\noindent
The answer is provided by the general idea in integrability theory:
the one of equivalent hierarchies \cite{eqhier}.

\subsection{Low-triangular rotation
\label{ltr}}

In application to our situation, this idea implies
a simple inverse procedure  to construct all the
matrices which can be expressed through their eigenvalues,
with the skew $\tau$-functions (upper-right corner minors)
described by a counterpart of (\ref{tau1m}).

Consider the first fundamental representation.
We begin with the simple matrix $\tilde\Lambda$ in (\ref{Ttp})
which is diagonalized by the Vandermonde rotation and, hence,
the matrix elements of $e^{\tilde\Lambda}$ are explicitly
expressed through its eigenvalues.
Moreover, these elements are given by simple formulas like (\ref{teLcorner}).

\bigskip

Now consider an additional rotation of $\tilde\Lambda$
by another
matrix $U$.
For a given $U$, the new  matrix $\Lambda_U=U\tilde\Lambda U^{-1}$ is
still explicitly expressed through its eigenvalues.
Moreover, when $U$ is lower triangular,
the corner matrix element is just multiplied by $U_{11}U_{NN}^{-1}$:
\be
\Big(e^{\Lambda_U}\Big)_{1N} =\Big(U_{11}U_{NN}^{-1}\Big)\cdot\sum_k^\infty
{{\cal S}_{[k]}(\lambda)\over (k+r)!}
\label{teUcorner}
\ee
preserving its simple structure.

In fact this is {\bf a characteristic property of the skew $\tau$-functions},
i.e. the minors of size $k$ at the upper right corner:
{\bf they transform in a simple way under conjugation by lower triangular matrices} $U$.
Indeed, such minors are multiplied from the left and from the right by the sub-matrices $U_{ab}$ and
$U^{-1}_{N+1-a,N+1-b}$ with $a,n\leq k$, which are also low triangular, thus, the determinant
is just multiplied by a product of diagonal elements:
\be
\tau^{(k)}_- (Ue^\Lambda U^{-1}) = \left(\prod_{a=1}^k {U_{aa}}{U^{-1}_{N+1-a,N+1-a}}\right)\cdot
\tau^{(k)}_- (e^\Lambda) \ \ \ \ \ \ \hbox{for low-triangular}\  U
\ee
Likewise, the minor at the lower left corner,
which we name "double-skew" $\tau$-functions
$\tau_{=}$ and consider in the next section \ref{doubleskew},
are transformed in a simple way under conjugation with the upper-triangular matrices.

Thus, it is clear that the skew $\tau$-functions for connections $\Lambda$,
which differ from the Vandermonde-diagonalizable
(\ref{Ttp}) by a lower-triangular conjugation,
will all possess the eigenvalue representation (\ref{tau1m})
modulo simple factors made from the diagonal elements $U_{aa}$.
Therefore, this is not a big surprise that the seemingly complicated matrices $U$
in (\ref{Tps2})-(\ref{Tps4}) differ from the Vandermonde matrices just by low-triangular factors.

For example, in the case of matrices (\ref{Tps3}) and (\ref{Tps4}) with $N=3$ and $N=4$
the relevant rotations are
\be
\nn \\
\underbrace{\left(\begin{array}{ccc} 4 & 4 & 4 \\ \\
2\sqrt{2}\lambda_1&2\sqrt{2}\lambda_2&2\sqrt{2}\lambda_3 \\ \\
\lambda_1^2+ \lambda_2\lambda_3 &\lambda_2^2+ \lambda_1\lambda_3 & \lambda_3^2+ \lambda_1\lambda_2
\end{array}\right)}_V =
\underbrace{\left(
\begin{array}{cccc}
4&0&&0\\
&&&\\
0&2\sqrt{2}&&0\\
&&&\\
  -S_{[2]}(\lambda)&0&&2\\
 \end{array}
\right)}_U
\underbrace{\left(\begin{array}{ccc }
1&1&1 \\ \\ \lambda_1&\lambda_2&\lambda_3  \\ \\
  \lambda_1^2&\lambda_2^2&\lambda_3^2
\end{array}\right)}_{\text{Vandermonde}}
\\ \nn
\ee
and
{\footnotesize
\be
\nn \\
\!\!\!\!\!\!\!\!\!\!\!\!\!\!
\underbrace{\left(\begin{array}{ccc} 60 &\ldots\!\!\! \\ \\ 20\sqrt{3}\lambda_1&&  \\ \\
\frac{\sqrt{3}}{2}\Big(20\lambda_1^2 -3(\lambda_1^2+\lambda_2^2+\lambda_3^2+\lambda_4^2)\Big)
& \\ \\
4\lambda_1^3- \lambda_1(\lambda_2^2+\lambda_3^2+\lambda_4^2) - 5\lambda_2\lambda_3\lambda_4
&
\end{array}\right)}_V =
\underbrace{\left(
\begin{array}{ccccc}
60&0&0& &0\\
&&&\\
0&20\sqrt{3}&0&&0\\
&&&\\
{-3\sqrt{3} } S_{[2]}(\lambda)&0&10\sqrt{3}&&0\\
&&&\\
-{5} S_{[3]}(\lambda)&-{7 } S_{[2]}(\lambda)&0&&10
\end{array}
\right)}_U
\underbrace{\left(\begin{array}{ccccc}
1&\ldots\!\! \\ \\ \lambda_1&  \\ \\
  \lambda_1^2&  \\ \\
  \lambda_1^3&
\end{array}\right)}_{\text{Vandermonde}}
\\ \nn
\ee}
Note that ${\cal S}_{[1]} = \sum_{i=1}^N \lambda_i = 0$, therefore
${\cal S}_{[2]} = \frac{1}{2}\sum_{i=1}^N \lambda_i^2$ and
${\cal S}_{[3]} = \frac{1}{3}\sum_{i=1}^N \lambda_i^3$.
Clearly, the ratios of diagonal elements of $U$ reproduce the needed boxed factors in
(\ref{Vma3}) and (\ref{Vma4}).

\section{Double skew $\tau$-functions and Toda recursion
\label{doubleskew}}

Now we will look at the opposite corner of the exponentiated matrix, this is what we call double-skew $\tau$-function $\tau_{=}$.
Since our connections are not symmetric, this is quite a different quantity
with different properties.

\subsection{Low-triangular connection}

In fact,  $\tau_{=}$ is non-trivial already when there is no $T_+$ in the constant
connection $\Lambda$ in (\ref{Tps}), we emphasize this by adding a subscript $0$ to $\Lambda$.
Despite all the remaining generators commute and  exponential is nicely factorized,
\be
e^{\Lambda_0} = \prod_{s=1}^r e^{w_sT_-^{(s)}}
\label{Lambda0}
\ee
the matrix element
\be
\tau_{=}^{(k)} = \ <-\text{hw}_k|\ e^\Lambda_0\ |\text{hw}_k>\
= \sum_{i_1,\ldots,i_s} \left<-\text{hw}_k\left|\prod_{s=1}^r\frac{w_s^{i_s}\Big(T_-^{(s)}\Big)^{i_s}}{i_s!}
\right|\text{hw}_k\right>
\ee
is somewhat complicated.
The sum is restricted by the obvious condition
\be
\sum_{s=1}^r s\cdot i_s = \delta_k=k(N-k)
\ee
thus, it is just a finite degree polynomial in $\{w_s\}$, actually, a Schur polynomial.
Indeed,
in the first fundamental representation, the matrix element is independent of $\{i_s\}$:
\be
\left<-\text{hw}_1\left|\prod_{s=1}^r\Big(T_-^{(s)}\Big)^{i_s}\right|\text{hw}_1\right> \ =
\left(\prod_{i=1}^r \sigma_i = r!\right)\delta\left(\sum_{s=1}^r s\cdot i_s - r\right)
\label{prodsig}
\ee
thus,
\be
\tau_{=}^{(1)}(e^{\Lambda_0}) = r! \sum_{i_1,\ldots,i_s} \left(\prod_{s=1}^r  \frac{w_s^{i_s}}{i_s!}\right)
\delta\left(\sum_s s\cdot i_s-r\right) = \nn\\
= r! \oint\frac{dz}{z^{r+1}}\exp\left(\sum_{s=1}^r w_sz^s\right)
= r! \oint\frac{dz}{z^{r+1}} \left(\sum_{s=1}^r z^s{\cal S}_{[s]}\{w_k\}\right)
= r!\cdot{\cal S}_{[r]}\{w_k\}
\ee
where the Schur polynomial is now expressed
through the time variables $p_k=kw_k$
rather than through Miwa variables $\lambda_i$ as in the previous section \ref{skewtaus}), where the counterpart of the time variables $p_k$ in Schur polynomials was given by (\ref{logdet}).
In fact this follows directly from (\ref{Lambda0}), if one remembers that
$T_-^{(s)}$ act in the first fundamental representation as powers of $T_{-}$.

Dependence on representation is now described in a usual simple way:
the double-skew $\tau$-functions in the next fundamental representations
are iteratively provided by the Toda recursion:
\be
\tau_{=}^{(k+1)}(e^{\Lambda_0}) = \frac{\sigma_{k-1}^2}{ \sigma_k^2\cdot    \tau_=^{(k-1)}} \cdot
 \left\{ \tau_=^{(k)}\frac{\p \tau_=^{(k)}}{\p w_1^2} -
\left(\frac{\p \tau_=^{(k)}}{\p w_1}\right)^2\right\}
\label{tau=rec}
\ee
where the extra $\sigma$-factors $\sigma_k^2=k(r+1-k)$
account for the deviation of matrix elements
between non-highest/lowest states from (\ref{prodsig}):
\be
\left<r-b\left|\prod_{s=1}^r \Big(T_-^{(s)}\Big)^{i_s}\right|a\right> \ =
\left(\prod_{i=a+1}^{r-b} \sigma_i = \frac{r!}{\prod_{i=1}^{a}\sigma_i\prod_{j=1}^b \sigma_{ j}}\right)
\delta\left(\sum_{s=1}^r s\cdot i_s - (r-a-b)\right)
\ee
where we also used the symmetry $\sigma_{r+1-j}=\sigma_j$.
Note that all derivatives are with respect to $w_1$: this is the Toda chain, not Toda lattice recursion.
From this recursion one immediately obtains
\be
\tau_{=}^{(1)} = r!\cdot {\cal S}_{[r]}, \ \ \
\tau_{=}^{(2)} = r! (r-1)!\cdot {\cal S}_{[r-1,r-1]}, \ \ \
\tau_{=}^{(3)} = r!(r-1)!(r-2)!
\cdot {\cal S}_{[r-2,r-2,r-2]}, \ \ \ \ldots \ \ \nn \\
\boxed{
\tau_{=}^{(k)}\left(e^{\Lambda_0}\right)=
\frac{{\cal S}_{[(N-k)^k]}}{ d_{[(N-k)^k]}}
= \prod_{i=1}^k(N-i)! \cdot {\cal S}_{[\underbrace{N-k,\ldots,N-k}_{k \text{\ times}}]}\{w_s\}
}
\label{dskewtauSchur}
\ee
which follows from (\ref{tau=rec}) and
\be
\p_{p_1} {\cal S}_{[i_1,i_2,\ldots]} =
\sum_{s:\ i_s>i_{s+1}} {\cal S}_{[i_1,i_2,\ldots,i_{s-1}, i_s-1,i_{s+1},\ldots]}
\ee
where contributing are only terms with $i_s\neq i_{s-1}$.

Other elements of the triangular matrix $e^{\Lambda_0}$ are also expressed through the Schur functions:
\be
\!\!\!\!\!\!\!\!\!
e^{\sum_s w_s T^{(s)}_-} = \left(\begin{array}{ccccccc}
1&0&0&0&\ldots &0&0 \\ \\
\sigma_1{\cal S}_{[1]} &1&0&0&&0&0\\ \\
\sigma_1\sigma_2{\cal S}_{[2]} & \sigma_2{\cal S}_{[1]} & 1 & 0 && 0&0 \\ \\
\sigma_1\sigma_2\sigma_3{\cal S}_{[3]} & \sigma_2\sigma_3{\cal S}_{[2]} & \sigma_3{\cal S}_{[1]} & 1  &&0& 0
\\ \\
\ldots \\ \\
\sigma_1\ldots\sigma_{r-1}{\cal S}_{[r-1]} &\sigma_2\ldots\sigma_{r-1}{\cal S}_{[r-2]}&
\sigma_3\ldots\sigma_{r-1} {\cal S}_{[r-3]} & \sigma_4\ldots \sigma_{r-1}{\cal S}_{[r-4]}& & 1&0 \\ \\
\sigma_1\ldots\sigma_{r}{\cal S}_{[r]} &\sigma_2\ldots\sigma_{r}{\cal S}_{[r-1]}&
\sigma_3\ldots\sigma_{r} {\cal S}_{[r-2]} & \sigma_4\ldots \sigma_{r}{\cal S}_{[r-3]}&
\ldots& \sigma_r{\cal S}_{[1]}&1
\end{array}\right)
\label{expL0}
\ee
Then, the minors at the low left corner are
\be
\tau^{(1)}_= = \sigma_1\ldots \sigma_r\cdot {\cal S}_{[r]} = r!\cdot {\cal S}_{[r]}, \nn \\
\tau^{(2)}_= = \sigma_1\Big(\sigma_2\ldots\sigma_{r-1}\Big)^2 \sigma_r\cdot
 \Big({\cal S}_{[r-1]}^2-{\cal S}_{[r]} {\cal S}_{[r-2]}\Big)
= r!(r-1)!\left\{\left(\frac{\p {\cal S}_{[r]}}{\p p_1}\right)^2
-{\cal S}_{[r]}\frac{\p^2 {\cal S}_{[r]}}{\p p_1^2}\right\}
= r!(r-1)!\cdot {\cal S}_{[r-1,r-1]}, \nn \\
\ldots
\ee

\subsection{Back to (\ref{Tps})}

If one wants  to return from the simplified pure triangular connection $\Lambda_0$ to more
general ones, it is needed to convert expressions like
\be
\exp\left(\sum_{i=1}^r t_i T_+^{(i)} + \sum_{j=1}^r\bar t_j T_-^{(j)}\right)
\ee
into the "normal ordered" form
\be
\exp\left(\sum_{i=1}^r t_i T_+^{(i)} \right) \ G \ \exp\left(\sum_{j=1}^r\bar t_j T_-^{(j)}\right)
= e^H \  \tilde G \ e^{\bar H}
\ee
In general, for this one needs to apply the Campbell-Hausdorff formula \cite{CH}
{\footnotesize
\be
e^A e^B = \exp\left\{\int_{0}^1 dt \sum_{n=1}^\infty \frac{(-)^n}{n}
\left( e^{t\cdot ad_A}e^{t\cdot ad_B} -1\right)^{n-1} e^{t\cdot ad_A} \Big(A+B\Big)\right\}
= \nn \\
= \exp\left\{A+B + \frac{1}{2!}[A,B] + \frac{1}{2\cdot 3!}\Big([A,[A,B]]+[[A,B],B]\Big)
+ \frac{1}{2\cdot 4!}\Big(\left[[A,[A,B]],B\right] + \left[A,[[A,B],B]\right]\Big) + \ldots
\right\}
\ee}
\noindent
where $ad_A = [A, \ldots]$.
However, if one switches on a single $T_+$ with coefficient $u$ to get (\ref{Tps}),
the expansion in powers of $u$ is relatively simple:
\be
e^{uT_++\Lambda_0} = e^{\Lambda_0} + u \int_0^1 e^{t\Lambda_0} T_+e^{(1-t)\Lambda_0} dt+ \ldots +\nn\\
+ u^m \int_{0\leq t_1\leq\ldots\leq t_m<1} e^{t_1\Lambda_0}T_+ e^{(t_2-t_1)\Lambda_0}T_+\ \ldots\
T_+e^{(t_m-t_{m-1})\Lambda_0} T_+e^{(1-t_m)\Lambda_0} dt_1\ldots dt_m +\ldots
\ee
and the $u$-linear correction to $\tau^{(1)}_=$ is
\be
u \cdot r! \int_0^1 dt \sum_{i=1}^r\sigma_1^2 {\cal S}_{[N-i]}\{tw_s\}{\cal S}_{[i]}\big\{(1-t)w_s\big\}
\ee
and so on.
When all $w_{k\geq 2}=0$, the only non-vanishing time is $p_1$,
and, since ${\cal S}_R\{p_1\} = d_R p_1^{|R|}$ we return to the familiar result from
section \ref{elem}:
\be
\tau^{(k)}_=\left(e^{uT_++w_1T_-}\right)  = \left(\sqrt{\frac{w_1}{u}}\cdot\sinh(\sqrt{uw_1})\right)^{k(N-k)}
\ee

\bigskip

Now the question is {\bf what happens to Toda structure at all non-zero $w_k$, when $u\neq 0$?}

\bigskip

When $u=0$ we had a differential operator, which acted inside (\ref{expL0}):
\be
\begin{array}{|cccccc|}
\hline
&&&&&\ \ \ \ \ \ \ \ \ \ \\
\ldots &&&&&\\
&&&&&\\
\p_1 \uparrow\ \ &&\p_1\uparrow \ \ &&& \\
&&&&&\\
{\cal S}_{[r-2]} & \stackrel{\p_1}{\longrightarrow} & {\cal S}_{[r-3]} &    \stackrel{\p_1}{\longrightarrow} &\ldots& \\
&&&&&\\
\p_1 \uparrow\ \ &&\p_1\uparrow\ \ &&& \\
&&&&&\\
{\cal S}_{[r-1]} & \stackrel{\p_1}{\longrightarrow} & {\cal S}_{[r-2]} &    \stackrel{\p_1}{\longrightarrow} &\ldots& \\
&&&&&\\
\p_1 \uparrow \ \ &&\p_1\uparrow\ \ & &&\\
&&&&&\\
{\cal S}_{[r]} & \stackrel{\p_1}{\longrightarrow} & {\cal S}_{[r-1]} &    \stackrel{\p_1}{\longrightarrow} &\ldots& \\
&&&&&\\
\hline
\end{array}
\ee

\bigskip

Remarkably, something survives of this structure when $u\neq 0$. We will denote the corresponding matrix elements ($u$-deformed Schur polynomials) $\MS^{(N)}_{a,b}$, and, for the sake of brevity, the first column will be denoted as $\MS^{(N)}_{[i]}$, while the second one $\tilde\MS^{(N)}_{[i]}$.

As the simplest example,   there is  an $N$-independent differential operator,
\be
{\cal D}_1 = \underline{\p_1} + \underline{\frac{u}{6}\Big(p_1\p_1}+ 2p_2\p_2+3p_3\p_3 + \ldots\Big)
\underline{- \frac{u^2}{180}\Big(7p_1^2\p_1}+30p_2p_1\p_2+ (48 p_3p_1+24 p_2^2 ) \p_3+ \ldots \Big) +\nn\\
+\underline{\frac{u^3}{7560}\Big({71}p_1^3\p_1}+(-336p_2^2+436p_2p_1^2)\p_2
+(-2142p_3p_2+663p_3p_1^2+675p_2^2p_1)\p_3 +\ldots\Big)
+ \ldots
\ee
which connects $u$-deformed Schur polynomials $\MS^{(N)}_{[r]}$
at the low-left corner with the next ones, and $\tilde\MS^{(N)}_{[r-1]}=\MS^{(N)}_{[r-1]}$:
\be
\begin{array}{|cccccc}
\ldots \\
&&&&&\\
{\MS}_{[r-2]}^{(N)}\{u|p\} &  \\
&&&&&\\
? \uparrow \ \ &&\\
&&&&&\\
{\MS}_{[r-1]}^{(N)}\{u|p\} &
\stackrel{?}{\longrightarrow} & {\tilde \MS}_{[r-2]}^{(N)} &    \\
&&&&&\\
{\cal D}_1 \uparrow \ \   && ? \uparrow\ \ & &&\\
&&&&&\\
{\MS}_{[r]}^{(N)}\{u|p\} & \stackrel{{\cal D}_1}{\longrightarrow} & {\MS}_{[r-1]}^{(N)}\{u|p\} &    \ldots \\
&&&&&\\
\hline
\end{array}
\ee

It is easy to understand that such an operator, if exists at all, should contain rather
strange coefficients: in the case of $N=2$ it should connect the two explicitly known functions,
\be
{\cal D}_1\left( \sqrt{\frac{p_1}{u}}\sinh(\sqrt{up_1})\right) = \cosh(\sqrt{up_1})
\ee
It follows that the underlined terms are actually
\be
{\cal D}_1 = \underbrace{\frac{2}{1+\frac{\tanh(\sqrt{up_1})}{\sqrt{up_1}}}}_{
%\!\!\!\!\!\!\!\!\!\!\!\!\!\!\!\!
1+\frac{up_1}{6} - \frac{7(up_1)^2}{180} +\frac{71(up_1)^3}{7560}  -\frac{517(up_1)^4}{226800}
+ \frac{307(up_1)^5}{554400}- \ldots}
\!\!\!\!\!\!\!\!\!\!\!\!\!\!\!\!\!\!\!\!\!\!\!\!\!\!\!\!\!\!\!\!\!\!\!\!\!\!
\cdot \p_1 + O(\p_2,\p_3,\ldots)
\ee
The same example is sufficient to understand that this operator can not act deeper inside
the table, because
\be
{\cal D}_1  \cosh(\sqrt{up_1}) \ \neq \ \sqrt{\frac{u}{p_1}}\sinh(\sqrt{up_1})
\ee
Occasionally, in this particular case all the three matrix elements are related by a simple dilatation:
\be
{\cal D} = \frac{1}{p_1}\Big(p_1\p_1 + u\p_u\Big): \ \ \ \ \ \
 \sqrt{\frac{p_1}{u}}\sinh(\sqrt{up_1}) \ \stackrel{{\cal D}}{\longrightarrow} \ \cosh(\sqrt{up_1})
\ \stackrel{{\cal D}}{\longrightarrow} \ \sqrt{\frac{u}{p_1}}\sinh(\sqrt{up_1})
\ee
but this does not immediately generalize to $N>2$.

\subsection{Evaluation of $\MS^{(N)}_{[0]}$}

For reference, we list in the Appendix the first $u$-deformed Schur functions, i.e. the first few matrix elements
\be
\MS_{N-a,b}^{(N)} = \left(\prod_{j=b}^{N-a} {\sigma_j}\right)^{-1}
\left\{\exp\left(uT_++\sum_{s=1}^r \frac{p_s}{s}T_-^{(s)}\right)\right\}_{N-a,b}
\ee
This matrix has the symmetry $a\leftrightarrow b$,
but for $u\neq 0$ its elements change along the subdiagonals.
Also, $u$-corrections in $\MS_{[k]}^{(N)}={\cal S}_{[k]}+O(u)$ depend on $N$.

Here we discuss in detail only $\MS^{(N)}_{[0]}$, which is the element at the upper left or lower right corner of $e^\Lambda$,
i.e. it is the conventional $\tau$-function $\tau_1$.
$p_1$-derivatives convert it into elements in the first row and the last column,
while $u$-derivatives produce those in the first row and the first column,
i.e. $\MS^{(N)}_{[1]}$  etc.

The evaluation goes through the Gauss decomposition, only this time one needs
$e^\Lambda = e^LDe^U$, then $\MS_{[0]} = \Big(e^\Lambda\Big)_{11} = D_{11}$.
The triangular matrices $L$ and $U$ are made from all the $N^2-N$ roots (not only from
the highest weights $T^\pm_s$), while the diagonal one,
\be
D = \exp\left(\sum_{s=1}^r \alpha_s \cdot ad_{T^+}^s (T_s^-)\right)
\ee
can be thought to involve nothing else.
Moreover, dependence of the element $D_{11}$  on the matrix size is very simple:
\be
\MS_{[0]} = D_{11}^{(N)} = \exp\left( \sum_{s=1}^r \alpha_s
\underbrace{\prod_{j=1}^s\sigma_j^2}_{\frac{(N-1)!s!}{(N-s-1)!}}\right)
\ee
As to $\alpha_s$, it first appears for $N=s+1$, i.e. it can be evaluated from the value of
$\Big(e^\Lambda\Big)_{11}$ at $N=s+1$, provided all the lower $\alpha_{j<s}$ are already known.

To get some impression of what $\alpha_s$ looks like, we evaluate it in the case when
only $p_s$ is non-vanishing.
Then the relevant element of the $(s+1)\times (s+1)$ matrix is
\be
\Big(e^\Lambda\Big)_{11} = e^{(s!)^2\cdot \alpha_s}  =
\sum_{i=0}^\infty \frac{ \Big(s!(s-1)!u^sp_s\Big)^i}{\Big((s+1)i\Big)!}
\ee
and for the matrices of size $N\times N$ this should be raised to the power $\frac{(N-1)!}{s!(N-1-s)!}$.

In particular, when only $p_1\neq 0$ the only non-vanishing coefficient is $\alpha_1$,
and
\be
\MS^{(N)}_{[0]}(p_1) = \left(\sum_{i=0}^\infty \frac{(up_1)^i}{(2i)!}\right)^{N-1} =
\cosh^{r}(\sqrt{up_1})
\ee
familiar from s.\ref{elem}.
Similarly,
$$
\MS^{(N)}_{[0]}(p_2) = \left(\sum_{i=0}^\infty \frac{(2u^2p_2)^i}{(3i)!}\right)^{\frac{(N-1)(N-2)}{2!}}
\cdot\left(1 - \frac{(u^2p_2)^2}{4} + \frac{3\cdot 49(u^2p_2)^3}{6!}
 -\frac{4\cdot 27\cdot 163\cdot 269(u^2p_2)^4}{11!} + \ldots\right)^{\frac{(N-1)(N-2)(N-3)}{3!}}\cdot
 $$

\centerline{$
\cdot\left(1   + \frac{12\cdot 17(u^2p_2)^3}{7!}
 -\frac{64\cdot 81\cdot 7279(u^2p_2)^4}{11!} + \ldots\right)^{\frac{(N-1)(N-2)(N-3)(N-4)}{4!}}
 \cdot\left(1
 -\frac{3\cdot 47\cdot 7279(u^2p_2)^4}{16\cdot 7} + \ldots\right)^{\frac{(N-1)(N-2)(N-3)(N-4)(N-5)}{5!}}
 \ldots
$}

\be
\MS^{(N)}_{[0]}(p_3) = \left(\sum_{i=0}^\infty \frac{(12u^3p_3)^i}{(4i)!}\right)^{\frac{(N-1)(N-2)(N-3)}{3!}}\cdot
\ldots \nn \\
\ldots
\ee

\bigskip

For $N=3$ and $p_1=0$ the entire matrix is
\be
\left(\begin{array}{ccccc}
\sum_{i=0}^\infty \frac{(2u^2p_2)^i}{(3i)!} & \stackrel{p_2^{2/3} \p_{_2} p_2^{1/3}}{\longleftarrow}
&\sqrt{2}u \sum_{i=0}^\infty \frac{(2u^2p_2)^i}{(3i+1)!}
&\stackrel{p_2^{1/3} \p_{_2} p_2^{2/3}}{\longleftarrow}
& 2u^2 \sum_{i=0}^\infty \frac{(2u^2p_2)^i}{(3i+2)!}
\\ \\ \downarrow\, p_2\p_2  && \downarrow \, p_2^{2/3} \p_{_2} p_2^{1/3}
&& \downarrow \,p_2^{1/3} \p_{_2} p_2^{2/3}\\ \\
\frac{1}{\sqrt{2}u}\sum_{i=1}^\infty \frac{(2u^2p_2)^i}{(3i-1)!}
&\stackrel{  p_2\p_2  }{\longleftarrow}& \sum_{i=0}^\infty \frac{(2u^2p_2)^i}{(3i)!}
&\stackrel{p_2^{2/3} \p_{_2} p_2^{1/3} }{\longleftarrow}
&\sqrt{2}u \sum_{i=0}^\infty \frac{(2u^2p_2)^i}{(3i+1)!}
\\ \\ \downarrow \, p_2^{4/3} \p_{_2} p_2^{-1/3}
&& \downarrow \, p_2\p_2  && \downarrow \, p_2^{-1/3} \p_{_2} p_2^{1/3}  \\ \\
\frac{1}{2u^2}\sum_{i=1}^\infty \frac{(2u^2p_2)^i}{(3i-2)!}
&\stackrel{ p_2^{4/3} \p_{_2} p_2^{-1/3} }{\longleftarrow}
& \frac{1}{\sqrt{2}u}\sum_{i=1}^\infty \frac{(2u^2p_2)^i}{(3i-1)!}
&\stackrel{p_2\p_2}{\longleftarrow}&
\sum_{i=0}^\infty \frac{(2u^2p_2)^i}{(3i)!}
\end{array}\right)
\ee

\section{Discussion and conclusion\label{con}}

\subsection*{Summary}

In this paper, we considered peculiar matrices $e^{\Lambda}$,
which represent $P$-exponentials of flat connections in the case when
the connections are constant.
Moreover, following \cite{Go,JdB,ACI,PerGrav,HKP}, we concentrated on $r$-parametric
matrices of peculiar form (\ref{Tps}), which are fully defined
through their eigenvalues, though the explicit expression is somewhat
sophisticated, see (\ref{Tpsev5})-(\ref{Tpsev1}).
These matrices are highly asymmetric, still the minors at their corners
all possess interesting properties related to Toda integrable systems.
While the minors at the upper-left and lower-right corners are just the
usual Toda lattice $\tau$-functions considered in \cite{GKLMM},
which are equally well expressed in term of time and Miwa variables,
those at the upper-right and lower-left corners,
which we call respectively "skew" and "double-skew" $\tau$-functions
did not attract enough attention in the literature.

\begin{picture}(220,220)(-130,-10)
\put(50,98){\mbox{$e^\Lambda \ = \ e^{uT_++\sum_s w_sT^{(s)}_-}$}}
\put(0,0){\line(1,0){200}}
\put(0,0){\line(0,1){200}}
\put(200,0){\line(0,1){200}}
\put(0,200){\line(1,0){200}}
\put(50,0){\line(0,1){50}}
\put(150,0){\line(0,1){50}}
\put(50,200){\line(0,-1){50}}
\put(150,200){\line(0,-1){50}}
\put(0,50){\line(1,0){50}}
\put(0,150){\line(1,0){50}}
\put(200,50){\line(-1,0){50}}
\put(200,150){\line(-1,0){50}}
\put(20,173){\mbox{$\tau^{(k)}$}}
\put(10,23){\mbox{$\tau^{(k)}_=\{w\}$}}
\put(160,173){\mbox{$\tau^{(k)}_-(\lambda)$}}
\put(170,23){\mbox{$\bar\tau^{(k)}$}}
\end{picture}

\noindent
We demonstrated that the skew $\tau$-functions are best studied in Miwa-like (eigenvalue)
$\lambda$-variables, while the double skew ones in terms of the ordinary
time variables $t_k=w_k$.
In all the four cases, relations between the minors of different sizes
are  provided by very similar Toda-like equations reflecting existence of
the Pl\"ucker relations, but in a somewhat different way.
However, explicit expressions for $w_s$ through $\lambda$, needed in the case of $\tau_-$
and for matrix elements trough $w_s$ and $u$ in the case of $\tau_=$
remain to be found and understood.

\subsection*{Speculations
\label{spec}}

In fact, until recently, most attention in this field was concentrated around
the classical approximation, where the AdS/CFT correspondence is straightforward
and the non-Abelian nature of Chern-Simons theory plays almost no role.
In result, the observables were the limits of large spin
representations, which factorize and reduce to quantum dimensions for the closed Wilson lines.

However, in fact, the relevant observables in $3d$ higher spin theory
have to be the full-fledged knot invariants, which are far less trivial quantities,
made from quantum Racah and mixing matrices \cite{MMMkn2}.
Fortunately, in the simplest cases, which are in fact physically relevant
they are already known, and one can start extracting physical information from this knowledge.

For example, the Hopf link is described
in the case of two symmetric representations $[r]$ and $[s]$,
 by a wonderful hypergeometric series \cite{Arthalinks}:
\be
H_{r,s}(q,N) = D_rD_s\left(1 + \sum_{k=1}^{{\rm min}(r,s)}\Big(q-q^{-1}\Big)^k(-)^k
q^{k(k+3)/2-k(r+s+N)}
\prod_{j=0}^{k-1}
\frac{[r-j]![s-j]!}{[j+N]!}\right)
\label{Hopf}
\ee
where $n=\frac{q^n-q^{-n}}{q-q^{-1}}$
is the quantum number and $D_r = \frac{[N+r-1]!}{[r]!}$ is the quantum dimension of
representation $[r]$.
In ${\rm AdS}_3$ studies, one picks up only the leading factor $D_rD_s$
and interprets its logarithm as the entanglement and thermal entropy,
depending on the choice and interpretation of parameters $r$ and $s$.
One option is to relate $r$ to the mass of the "background" black hole,
while $s$ to that of the probe one.
Another option is to relate $r$ to the length of
a segment in calculation of the entanglement entropy.
What still remains obscure is a proper interpretation of the fixed-gauge open Wilson lines in terms of knot theory.

The real question is, however, to clarify the role of $q$ and interpret the
whole sum in (\ref{Hopf}) in terms of the higher spin theory.
This subject is closely related to the role of
quantum hypergeometric functions and operator-valued $\tau$-functions in
conformal theory.
We hope to elaborate on these issues in future publications.

\subsection*{Analytic continuation in $N$}

As we established in s.3.6, the skew $\tau$-functions are associated with the Toda molecule hierarchy, which is determined by conditions (\ref{Todam}). In fact, one can go even further to check that
\be\label{Todam2}
\tau_-^{(0)}=1,\ \ \ \ \ \tau_-^{(N)}=1
\ee
due to constraint (\ref{al}). One can immediately consider the limit of $N\to\infty$ just by lifting the second part of these conditions (\ref{Todam2}) and remaining with a generic Toda chain forced hierarchy \cite{KMMOZ}. However, an analytic continuation to arbitrary (non-integer) $N$ is hard, because of the first constraint $\tau_-^{(0)}=1$ and of the Toda chain recursion which connects three $\tau_-^{(k)}$ at neighbour values. This is, however, not necessary for higher spin theory: one only needs an (infinite) algebra\footnote{This algebra emerges as the universal enveloping algebra $U(sl(2,R))$ at the fixed value of the second Casimir equal to $(\mu^2-1)/4$.} hs[$\mu$] parameterized by a continuous parameter $\mu$ that at a generic value of $\mu$ contains the only finite subalgebra $sl(2)$, but at $\mu=N$ it factorizes into $sl(N)$ and an (infinite) ideal \cite{hsa}. In terms of the Toda forced hierarchy, this means that, at generic $\mu$, one considers solutions to the complete forced hierarchy parameterized by $\mu$: $\tau^{(k)}_-(\mu)$. One can define $\mu$ as follows: consider $\tau^{(k)}_-(\mu)$ as a function of parameter $k$ analytically continued to arbitrary values of $k$. Then, define $\mu$ as the location of a zero of $\log\tau_-^{(k)}$, i.e. $\tau^{(k)}_-(\mu)|_{k=\mu}=1$.
The presence of the $sl(2)$ subgroup is related to the condition of the forced hierarchy $\tau_-^{(0)}=1$, while at integer values of $\mu=N$ there emerges an additional condition $\tau_-^{(N)}=1$. We are planning to return to the manifest description of this generic solution $\tau_-^{(k)}(\mu)$ elsewhere.

\subsection*{Conclusion}

This paper is largely motivated by the recent results in \cite{Go,JdB,ACI,PerGrav,HKP}.
As we tried to argue, they almost embed the studies of open Wilson lines
in 3d higher spin theory into the general context of integrability theory.
Of crucial importance is the very interest to {\it open} Wilson lines \cite{JdB,ACI},
which are not normally studied within the Chern-Simons context,
but which reveal a lot of structures not seen (remaining hidden)
in conventional theory of knot polynomials.
On the other hand, knot theory needs quantization to become really interesting,
and combination of these two ingredients, integrability and quantization,
can finally bring to light the old attempts \cite{GKLMM,MorVin,Mir,Khar,qtau} to build the
operator-valued $\tau$-functions for quantum groups.
This can become a meeting point of the three popular subjects of the last decade:

$\bullet$ $2d$ conformal + $3d$ knot theory,

$\bullet$ quasiclassics/Stokes theory + wall crossing + cluster varieties,

$\bullet$ quantum gravity and higher spin theory + AdS/CFT correspondence

\noindent
A lot of links are already found between these subjects,
and all actually involve ideas from integrability theory, where the notion
of $\tau$-function or, better, a "matrix model $\tau$-function" \cite{UFN23}
with additional requirements to the choice of the group element
plays a central role.
As we tried to demonstrate in this paper, these are exactly the features
that are getting revealed and attract attention in \cite{HKP} and its precursors,
and from this point of view the most important next step should be quantization
providing connection to representation theory of quantum groups,
where
a considerable progress was recently made to serve the needs of the
CFT and knot polynomial research.

Especially important for the future research is understanding of integrability properties of semiclassical conformal blocks: a new and fast developing branch of CFT \cite{Hij,Perl}. At the same time, some older subjects like finite W-algebras (see a review in \cite{Tjin}) seem to be directly connected to our story. We hope to return to these issues elsewhere.

\section*{Acknowledgements}

We are grateful to V.~P.~Nair and F.~Novaes for useful conversations.

\noindent
A.Mor. acknowledges the hospitality of IIP at Natal during the work on this project. D.Mel. is thankful to S.~Klevtsov and the Institute for Theoretical Physics at the University of Cologne for hospitality at its final stage.

\noindent
Our work is partly supported by RFBR grants 14-02-00627 (D.Mel.), 16-01-00291 (A.Mir.), 16-02-01021 (A.Mor.), by grant 15-31-20832-Mol-a-ved (A.Mor.), by joint grants 16-51-53034-GFEN, 15-51-50034-YaF, 15-51-52031-NSC-a  (A.Mir. and A.Mor.) and the Science without Borders project 400635/2012-7 supported by the Brazilian National Counsel for Scientific and Technological Development (CNPq).

\newpage

\section*{Appendix. The first $u$-deformed Schur functions $\MS$}

We list here the first few matrix elements
\be
\MS_{N-a,b}^{(N)} = \left(\prod_{j=b}^{N-a} {\sigma_j}\right)^{-1}
\left\{\exp\left(uT_++\sum_{s=1}^r \frac{p_s}{s}T_-^{(s)}\right)\right\}_{N-a,b}
\ee
This matrix has a symmetry $a\leftrightarrow b$,
but for $u\neq 0$ its elements change along the subdiagonals.
Also, $u$-corrections in $\MS_{[k]}^{(N)}={\cal S}_{[k]}+O(u)$ depend on $N$.

\bigskip

$\bullet$  $N=2$:
\be
\begin{array}{clcl}
& \tilde\MS_{[-1]}^{(2)}=\MS_{1,2}^{(2)}= &  \sqrt{\frac{u}{p_1}}\sinh(\sqrt{up_1}) =&
\ \ \ \  u+ \frac{u^2p_1}{6} + \frac{u^3p_1^2}{120} + \ldots \\   \\
 &\ ? \uparrow {\cal D} & \\
 \hline \\
& \MS_{[0]}^{(2)} = & \cosh(\sqrt{up_1})=&
1 + \frac{up_1}{2} + \frac{u^2p_1^2}{24} + \frac{u^3p_1^3}{720} + \ldots  \\
\\ &{\cal D}_1 \uparrow {\cal D} & \\ \\
\tau^{(1)}_= = &
\MS_{[1]}^{(2)} = &\sqrt{\frac{p_1}{u}}\sinh(\sqrt{up_1}) =&
p_1 + \frac{up_1^2}{6} + \frac{u^2p_1^3}{120} + \frac{u^3p_1^4}{5040} + \ldots  \\
\hline \\
\tau^{(2)}_= = &  \MS_{[0]}^{(2)}-\MS_{[1]}^{(2)} \tilde\MS_{[-1]}^{(2)}=& 1
\end{array}
\ee

\bigskip

$\bullet$  $N=3$:
\be
\begin{array}{clcllll}
&\underbrace{\tilde\MS_{2,2}^{(3)}}_{\neq \MS_{[0]}^{(3)}}& \underbrace{ 1 }_{{\cal S}_{[0]} }
&+2up_1
&+\frac{u^2p_2}{3}+\frac{2u^2p_1^2}{3}
&+                   \frac{2u^3p_2p_1}{15} +\frac{4u^3p_1^3}{45}
& +\ldots \\
\hline \\
& \MS_{[0]}^{(3)}=  & \underbrace{ 1 }_{{\cal S}_{[0]}}
& +up_1
&+\frac{u^2p_2}{3}+\frac{u^2p_1^2}{3}
&+    \frac{u^3p_2p_1}{10} +\frac{2u^3p_1^3}{45}
& + \ldots \\
&? \uparrow \ \ & \\ \\
& \MS_{[1]}^{(3)} = &       \underbrace{ p_1 }_{{\cal S}_{[1]}}
&+ \frac{up_2}{2}+\frac{2up_1^2}{3}
&+ \frac{u^2p_2}{4}+\frac{2u^2p_1^2}{15}
&+ \frac{u^3p_2^2}{60} + \frac{2u^3p_2p_1^2}{45} + \frac{4u^3p_1^4}{315}
&  + \ldots  \\ \\
&{\cal D}_1 \uparrow \ \ & \\ \\
\tau^{(1)}_= = &
\MS_{[2]}^{(3)} = & \underbrace{ \frac{p_2}{2} +\frac{p_1^2}{2} }_{{\cal S}_{[2]}}
&+ \frac{up_2p_1}{3}+\frac{up_1^3}{6}
&+ \frac{u^2p_2^2}{24}+\frac{u^2p_2p_1^2}{12}+\frac{u^2p_1^4}{45}
&+ \frac{u^3p_2^2p_1}{90}+\frac{u^3p_2p_1^3}{105}+\frac{u^3p_1^5}{630}
&+ \ldots  \\   \hline  \\
\tau^{(2)}_= = &  \underbrace{\MS_{[1]}^{(3)}-\MS_{[2]}^{(3)} \tilde\MS_{2,2}^{(2)}}_{= \MS_{[1,1]}^{(3)}}=&
 \underbrace{ -\frac{p_2}{2}+\frac{p_1^2}{2}  }_{{\cal S}_{[11]}}
 & - \frac{up_2}{3}+\frac{up_1^3}{6}
& + \frac{u^2p_2^2}{24}  -\frac{u^2p_2p_1^2}{12}+\frac{u^2p_1^4}{45}
& + \frac{u^3p_2^2p_1}{90} - \frac{u^3p_2p_1^2}{105} + \frac{u^3p_1^5}{630}
 & + \ldots
\end{array}
\ee

\bigskip

$\bullet$  $N=4$:

\centerline{
$
\begin{array}{clcrlll}
&\underbrace{\tilde\MS_{3,2}^{(4)}}_{\neq \MS_{[1]}^{(4)}}& \underbrace{ p_1 }_{{\cal S}_{[1]} }
&  +\frac{3up_2}{2}+\frac{5up_1^2}{3}
& +\frac{u^2p_3}{2} +\frac{7u^2p_2p_1}{4}+\frac{91u^2p_1^3}{120}
&+\\ &&
&&  +\frac{7u^3p_3p_1}{20}
+\frac{3u^3p_2^2}{10}+\frac{57u^3p_2p_1^2}{80}+\frac{41u^3p_1^4}{252}
& +\ldots \\
\hline \\
& \MS_{[1]}^{(4)}=  & \underbrace{ p_1 }_{{\cal S}_{[1]}}
&  +up_2+\frac{7up_1^2}{6}
& +\frac{2u^2p_3}{3} +\frac{4u^2p_2p_1}{3}+\frac{61u^2p_1^3}{120}
&+\\ &&
&&  +\frac{13u^3p_3p_1}{30}
+\frac{u^3p_2^2}{5}+\frac{193u^3p_2p_1^2}{360} +\frac{547u^3p_1^4}{5040}
& + \ldots \\   \\
&? \uparrow \ \ & \\ \\
& \MS_{[2]}^{(4)} = &       \underbrace{\frac{p_2}{2} +\frac{p_1^2}{2} }_{{\cal S}_{[2]}}
&+ \frac{up_3 }{2}+\frac{13up_2p_1}{12}+\frac{5up_1^3}{12}
&  +\frac{5u^2p_3p_1}{12}+\frac{u^2p_2^2 }{4}+\frac{29p_2p_1^2}{48}+\frac{91u^2p_1^4}{720}
&+\\ &&
&  + \frac{3u^3p_3p_2}{20}-\frac{103u^3p_3p_1^2}{720}
&+\frac{23u^3p_2^2p_1}{120}+\frac{1573u^3p_2p_1^3}{10080}+\frac{41u^3p_1^5}{2016}
&  + \ldots  \\ \\
&{\cal D}_1 \uparrow \ \ & \\ \\
\tau^{(1)}_= = &
\MS_{[3]}^{(4)} = & \underbrace{ \frac{p_3}{3}+ \frac{p_2p_1}{2} +\frac{p_1^3}{6} }_{{\cal S}_{[3]}}
&+ \frac{up_3p_1}{3}+\frac{up_2^2}{6}+\frac{5p_2p_1^2}{12}+\frac{up_1^4}{12}
& +\frac{u^2p_3p_2}{6}+\frac{17u^2p_3p_1^2}{120}+\frac{11u^2p_2^2p_1}{60}+\frac{13u^2p_1^5}{720}
&+\\ &&
&  +\frac{u^3p_3^2}{30}-\frac{19u^3p_3p_2p_1}{180}+\frac{41u^3p_3p_1^3}{1260}
&+\frac{u^3p_2^3}{60}+\frac{331u^3p_2^2p_1^2}{5040}+\frac{53u^3p_2p_1^4}{2016}+\frac{41u^3p_1^6}{18144}
&+ \ldots  \\   \\  \hline \\
\tau^{(2)}_= = &  \underbrace{\MS_{[2]}^{(4)}-\MS_{[3]}^{(4)} \tilde\MS_{3,2}^{(4)}}_{= \MS_{[2,2]}^{(4)}}=&
 \underbrace{ -\frac{p_3p_1}{3}+\frac{p_2^2}{4}+\frac{p_1^4}{12}   }_{{\cal S}_{[22]}}
 & -\frac{7up_3p_1^2}{18}+\frac{up_2^2p_1}{6}+\frac{up_1^5}{18}
& +\frac{u^2p_3^2}{12}-\frac{u^2p_3p_1^3}{5}+\frac{u^2p_2^2p_1^2}{15}+\frac{u^2p_1^6}{60}
&+\\ &&
&  +\frac{u^3p_3^2p_1}{10}-\frac{u^3p_3p_2^2}{30}-\frac{223u^3p_3p_1^4}{3780}
&+\frac{11u^3p_2^2p_1^3}{630}+\frac{17u^3p_1^7}{5670}
 & + \ldots
\end{array}
$
}

\end{document}